\documentclass[11pt]{article}
\usepackage{jheppub}
\usepackage{amsmath,amssymb,amsfonts,graphicx,slashed,color,amsthm,mathtools,upgreek}

\newcommand\ack[1]{{\begin{color}{red} (#1)\end{color}}}

\newcommand{\bC}{\mathbb{C}}
\renewcommand{\a}{\alpha}

\newcommand{\cA}{\mathcal{A}}
\newcommand{\pa}{\partial}

\newcommand{\be}{\begin{equation}}
\newcommand{\ee}{\end{equation}}
\newcommand{\bea}{\begin{eqnarray}}
\newcommand{\eea}{\end{eqnarray}}
\newcommand{\beq}{\begin{equation}}
\newcommand{\eeq}{\end{equation}}
\newcommand{\beqn}{\begin{eqnarray}}
\newcommand{\eeqn}{\end{eqnarray}}

\newcommand{\bs}{\boldsymbol}

\newcommand{\vx}{\mathbf{x}}

\newcommand{\cL}{\mathcal{L}}
\newcommand{\cH}{\mathcal{H}}
\newcommand{\cS}{\mathcal{S}}

\newcommand{\Tr}{\mathrm{Tr}}

\usepackage{upgreek}

\theoremstyle{definition}

\title{Entanglement Entropy and the Colored Jones Polynomial}

\author[a, b]{Vijay Balasubramanian}
\author[a]{\!, Matthew DeCross}
\author[c]{\!, Jackson Fliss}
\author[a]{\!, Arjun Kar}
\author[c]{\!, Robert G. Leigh}
\author[a]{\!, Onkar Parrikar}

\affiliation[\,a]{David Rittenhouse Laboratory, University of Pennsylvania,\\
209 S.33rd Street, Philadelphia PA, 19104, U.S.A.}
\affiliation[\,b]{Theoretische Natuurkunde, Vrije Universiteit Brussel (VUB), and \\ International Solvay Institutes, Pleinlaan 2, B-1050 Brussels, Belgium.}

\affiliation[\,c]{Department of Physics, University of Illinois,\\
 1110 W. Green St., Urbana IL 61801-3080, U.S.A.}

\abstract{
We study the multi-party entanglement structure of states in Chern-Simons theory created by performing the path integral on 3-manifolds with linked torus boundaries, called link complements. For gauge group $SU(2)$, the wavefunctions of these states (in a particular basis) are the colored Jones polynomials of the corresponding links. We first review the case of $U(1)$ Chern-Simons theory where these are stabilizer states, a fact we use to re-derive an explicit formula for the entanglement entropy across a general link bipartition. We then present the following results for $SU(2)$ Chern-Simons theory: (i) The entanglement entropy for a bipartition of a link gives a lower bound on the genus of surfaces in the ambient $S^3$ separating the two sublinks. (ii) All torus links (namely, links which can be drawn on the surface of a torus)  have a GHZ-like entanglement structure -- i.e., partial traces leave a separable state.  By contrast, through explicit computation, we test in many examples that hyperbolic links (namely, links whose complements admit hyperbolic structures) have W-like entanglement -- i.e., partial traces leave a non-separable state. (iii) Finally, we consider hyperbolic links in the complexified $SL(2,\bC)$ Chern-Simons theory, which is closely related to 3d Einstein gravity with a negative cosmological constant. In the limit of small Newton constant, we discuss how the entanglement structure is controlled by the Neumann-Zagier potential on the moduli space of hyperbolic structures on the link complement.

}


\begin{document}

\maketitle

\parskip=10pt

\section{Introduction}
In simple quantum systems, such as collections of qubits, entanglement structure has been well-studied. The quintessential example of a two qubit entangled state is a Bell pair\footnote{The Hilbert space corresponding to a single qubit is identified with $\bC^2$, with an orthonormal basis typically labelled by $|0\rangle $ and $|1\rangle $. The Hilbert space corresponding to two qubits is $\bC^2\otimes \bC^2$, for three qubits is $\bC^2\otimes \bC^2 \otimes \bC^2$ and so on.}
\beq
|\psi_{\mathrm{Bell}}\rangle  = \frac{1}{\sqrt{2}} \Big(|0\rangle \otimes |0\rangle + |1\rangle \otimes |1\rangle \Big).
\eeq
If we trace out one of the qubits, then we are left with a \emph{mixed} state
\beq\label{rdm}
\rho_1 = \mathrm{Tr}_{2}\,|\psi_{\mathrm{Bell}}\rangle\langle \psi_{\mathrm{Bell}}| = \frac{1}{2} |0\rangle \langle 0| + \frac{1}{2} |1\rangle \langle 1|,
\eeq
which we should think of as an ensemble, or a probability distribution over pure quantum states in the one qubit Hilbert space. A good measure of the entanglement between the original two qubits is  the von Neumann entropy of the reduced density matrix \eqref{rdm}, also known as the \emph{entanglement entropy} 
\beq
S_{EE} = -\mathrm{Tr}_1\,\rho_1\ln\rho_1 = \ln\,(2). 
\eeq
The Bell state should be contrasted with states of the form $|0\rangle \otimes |0\rangle, |0\rangle \otimes |1\rangle$, etc., which are completely factorized and have no entanglement. The entanglement entropy thus measures the non-factorizability of a  state. 

For larger systems, one can construct states with more intricate patterns of entanglement. For instance, with three qubits one can construct the following two types of multi-party entangled states \cite{Dur:2000zz}:
\beq
|\psi_{\mathrm{GHZ}} \rangle = \frac{1}{\sqrt{2}} \Big(|000\rangle + |111\rangle \Big),
\label{GHZstate}
\eeq
\beq
|\psi_{\mathrm{W}} \rangle = \frac{1}{\sqrt{3}} \Big(|001\rangle + |010\rangle+|100\rangle \Big),
\label{Wstate}
\eeq
where we have neglected to write the tensor product symbols in favor of simpler notation. As we shall see, these two states carry different types of entanglement.
If we trace over one of the factors in the GHZ state, we get the reduced density matrix
\beq
\rho_{12} = \mathrm{Tr}_3 \,|\psi_{\mathrm{GHZ}}\rangle\langle \psi_{\mathrm{GHZ}}| = \frac{1}{2} |00\rangle \langle 0 0| + \frac{1}{2} |1 1\rangle \langle 11|.
\eeq
Thought of as a (mixed) two-qubit state on the first two qubits, $\rho_{12}$ is a classical probabilistic mixture over \emph{product} states, namely $|00\rangle $ and $|11\rangle$. In quantum information theory, such a state $\rho_{12}$ is called \emph{separable}. In other words, the reduced density matrix $\rho_{12}$ contains no quantum entanglement -- 
all the entanglement between qubit 1 and qubit 2 came from their  mutual relatonship with qubit 3 which was traced out.
 On the other hand, if we trace over one of the factors in the W state, we obtain the reduced density matrix
\beq
\tilde{\rho}_{12} = \mathrm{Tr}_3 \,|\psi_{\mathrm{W}}\rangle\langle \psi_{\mathrm{W}}| = \frac{1}{3} |00\rangle \langle 0 0| + \frac{2}{3} |\Psi_+\rangle \langle \Psi_+|, \quad |\Psi_+\rangle = \frac{|01\rangle + |10\rangle}{\sqrt{2}}.
\eeq
In this case, $\tilde{\rho}_{12}$ is once again a probabilistic mixture over two qubit states, namely $|00\rangle$ and $|\Psi_+\rangle$, but importantly $\Psi_+$ is \emph{not} a product state. In other words, the state $\tilde{\rho}_{12}$ contains quantum entanglement between qubit 1 and qubit 2; in this case we say that $\tilde{\rho}_{12} $ is not separable. In this sense, the quantum entanglement structure of the W-state is different from that of the GHZ state. 
Increasing the number of qubits increases the possible patterns of entanglement very quickly. In fact, for four or more qubits the SLOCC\footnote{SLOCC stands for stochastic local operations and classical communication. This classification effectively amounts to studying the equivalence classes of states in the full Hilbert space under a quotient by local actions of $SL(2,\bC)$, namely $\frac{\bC^2 \otimes \bC^2 \otimes \cdots \bC^2}{SL(2,\bC) \times SL(2,\bC) \times \cdots SL(2,\bC)}$. } classification gives classes of states some of which  contain continuous families  with fundamentally different patterns of entanglement \cite{2002PhRvA..65e2112V}. The situation is  going to be even richer for quantum field theories. 

\begin{figure}[t]
\centering
\includegraphics[height=3cm]{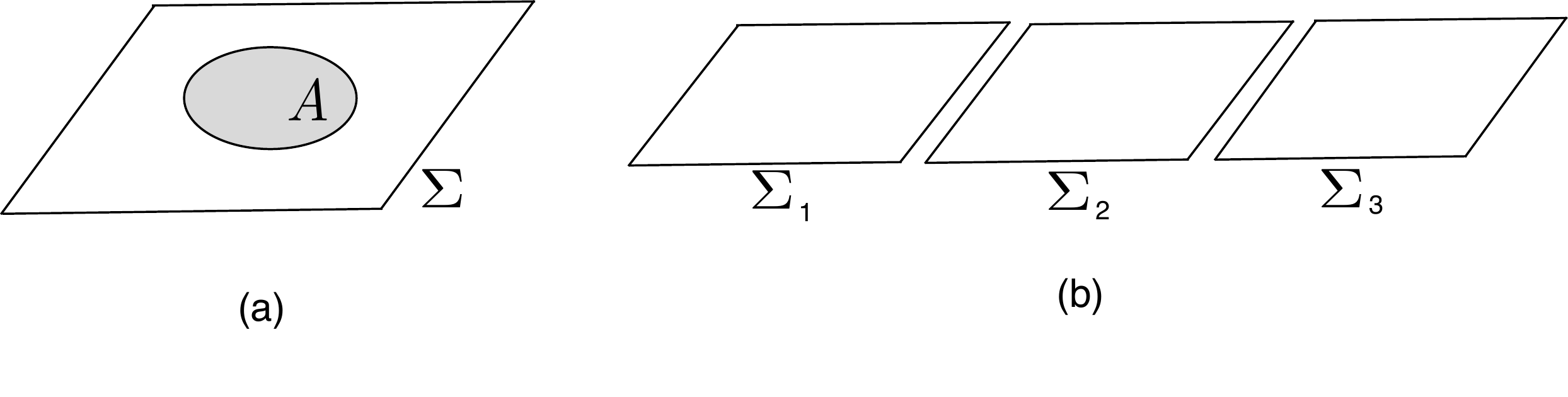}\label{fig1}
\caption{\small{\textsf{(a) The typical setup for studying entanglement entropy in quantum field theory involves choosing a connected spatial slice $\Sigma$ and partitioning it into two subregions $A$ (the shaded disc) and its complement $\bar{A}$. (b) In the present paper, we are interested in considering disconnected Cauchy surfaces $\Sigma = \Sigma_1 \cup \Sigma_2 \cup \cdots$ and studying the entanglement between these various disconnected components. }}}
\end{figure}

The typical setup for considering entanglement entropy in relativistic quantum field theories is as follows: one starts with a \emph{connected}, codimension-one, spacelike hypersurface, i.e., a Cauchy surface $\Sigma$. In quantum field theory, one associates a Hilbert space $\mathcal{H}(\Sigma)$ to such a surface. We pick some pure state $|\psi\rangle \in \mathcal{H}(\Sigma)$. Now let us imagine partitioning $\Sigma$ into two regions $A$ and its complement $\bar{A}$ (see Fig.~\ref{fig1}a). 
If the Hilbert space on $\Sigma$ factorizes as $\mathcal{H}(\Sigma) = \mathcal{H}(A) \otimes \mathcal{H}(\bar{A})$, then one can trace over one of the factors and obtain the reduced density matrix corresponding to $\psi$ on the subregion $A$:
\beq
\rho_A = \mathrm{Tr}_{\bar{A}} |\psi\rangle \langle \psi |. 
\eeq
Generically, the density matrix $\rho_A$ so obtained is mixed and the von Neumann entropy $S(A) = -\mathrm{Tr}_A\,(\rho_A\ln\rho_A)$ measures the entanglement entropy between the region $A$ and its complement. The entropy computed this way is typically divergent in the continuum limit, owing to the short-distance entanglement near the boundary between $A$ and $\bar{A}$, but these divergences are by now well-understood. 

In the present paper we will consider a different setup. Instead of considering a connected spatial slice, we will be interested in disconnected spatial slices of the form
\beq
\Sigma = \Sigma_1 \cup \Sigma_2 \cup  \cdots \cup \Sigma_n,
\eeq
such as the one shown in Fig.~\ref{fig1}(b). As a consequence, the Hilbert space $\mathcal{H}(\Sigma)$ naturally factorizes
\beq
\mathcal{H}(\Sigma) = \mathcal{H}(\Sigma_1) \otimes \mathcal{H}(\Sigma_2) \otimes\cdots \otimes \mathcal{H}(\Sigma_n) .
\eeq
We can then ask for the entanglement structure of states in $\mathcal{H}(\Sigma)$ with respect to this factorization. We will sometimes refer to this type of entanglement as \emph{multi-boundary entanglement}, in order to distinguish it from the other more conventional setting involving connected spatial slices. 

Multi-boundary entanglement was considered recently in the context of the AdS/CFT correspondence in \cite{Balasubramanian:2014hda, Marolf:2015vma} (see also \cite{Peach:2017npp}). In these papers, the conformal field theory (CFT) is 1+1 dimensional, and the Cauchy surface $\Sigma$ is a union of $n$ circles. Further, the states of interest are those dual to classical asymptotically-AdS multi-boundary wormhole geometries. The holographic entropies of entanglement between the various boundary circles can be studied using the Ryu-Takayanagi formula \cite{Ryu:2006bv}. Ideally, one would also like to perform similar entanglement computations entirely using field theory methods (i.e., without using the AdS dual); this was partly accomplished in \cite{Balasubramanian:2014hda, Marolf:2015vma} in certain special limits. Crucially, the  CFT states could  be obtained by performing the Euclidean field theory path integral on certain Riemann surfaces with $n$ circle boundaries.  At special points on the moduli space of these Riemann surfaces, the field theory computation became tractable. However, at a generic point on the moduli space, the computation is too difficult to perform explicitly. It is thus natural to look for a  ``simpler'' class of quantum field theories (as compared to CFTs), where we might be able to study multi-boundary entanglement using field theory techniques. A natural candidate is the class of topological quantum field theories (TQFTs) \cite{Witten:1988ze, Atiyah1988}. 

Motivated by this, some of the authors of the present paper explored multi-boundary entanglement in  Chern-Simons theory, in \cite{Balasubramanian:2016sro} (see also \cite{Salton:2016qpp}).\footnote{Chern Simons theory is also holographic, in the sense that it can be realized as the worldvolume theory of A-branes in topological string theory, and is dual to topological closed strings on 6d resolved conifold geometries.} The Cauchy surface $\Sigma$ was taken to be $n$ copies of a torus, and the states of interest were created by performing the path integral of Chern-Simons theory on {\it link complements} with $n$ torus boundaries. A link complement is a manifold obtained by removing a link from the 3-sphere (see Sec.~\ref{sec2} for  details). In fact, with a particular choice of basis for the torus Hilbert space, the wavefunctions of these states are precisely the expectation values of Wilson loop operators in Chern-Simons theory, often called \emph{colored link invariants}. For the gauge group $SU(2)$, these are precisely the \emph{colored Jones polynomials}, as was famously shown by Witten \cite{Witten:1988hf}. The central observation in \cite{Balasubramanian:2016sro} was that these states live in the $n$-fold tensor product of the torus Hilbert space, and as such it is natural to study the entanglement between the various factors (i.e., multi-boundary entanglement) in these states. In other words, the colored Jones polynomial assigns a \emph{quantum entanglement structure} to a link in the 3-sphere. 
Recently, the R\'{e}nyi entropies for a class of torus links called $T(2,2n)$ were also studied in detail in \cite{1711.06474} for general gauge groups.

In the present paper, we will further explore this quantum information theoretic approach to link topology.  In Sec.~\ref{sec2}, we will review the construction of \cite{Balasubramanian:2016sro}, and show how previous results on multi-boundary entanglement in $U(1)$ Chern-Simons theory may be rewritten from the point of view of stabilizer groups.  In Sec.~\ref{mgb}, we will prove that the entanglement entropy between any two sub-links of an arbitrary link gives a lower bound on the minimal-genus Heegaard splitting which separates the two sub-links.  In Sec.~\ref{sec4},  we show that  in $U(1)$ and $SU(2)$ Chern-Simons theory all {\it torus links} (which can be drawn on the surface of a torus), have a GHZ-like entanglement structure, in that partial traces lead to a separable state.  This provides a sharp quantum-information theoretic characterization of the colored Jones polynomial for torus links.   By explicit computation, we also show that many hyperbolic links (whose link complements admit a hyperbolic structure) have  W-like entanglement, in that partial traces do not lead to separable states.    In  Sec.~\ref{sec5}, we further study hyperbolic links in the complexified $SL(2,\bC)$ Chern-Simons theory, which is of interest because of its close connection to Einstein gravity with a negative cosmological constant.     In an asymptotic limit (where one of the levels $\sigma \to \infty$, corresponding to small Newton constant) we discuss how the entanglement structure is controlled by the \emph{Neumann-Zagier} potential on the moduli space of (generically incomplete) hyperbolic structures on the link complement. 


\section{Setup} \label{sec2}

\subsection{Link Complements and the Colored Jones Polynomial}

\begin{figure}
\centering
\includegraphics[height=4cm]{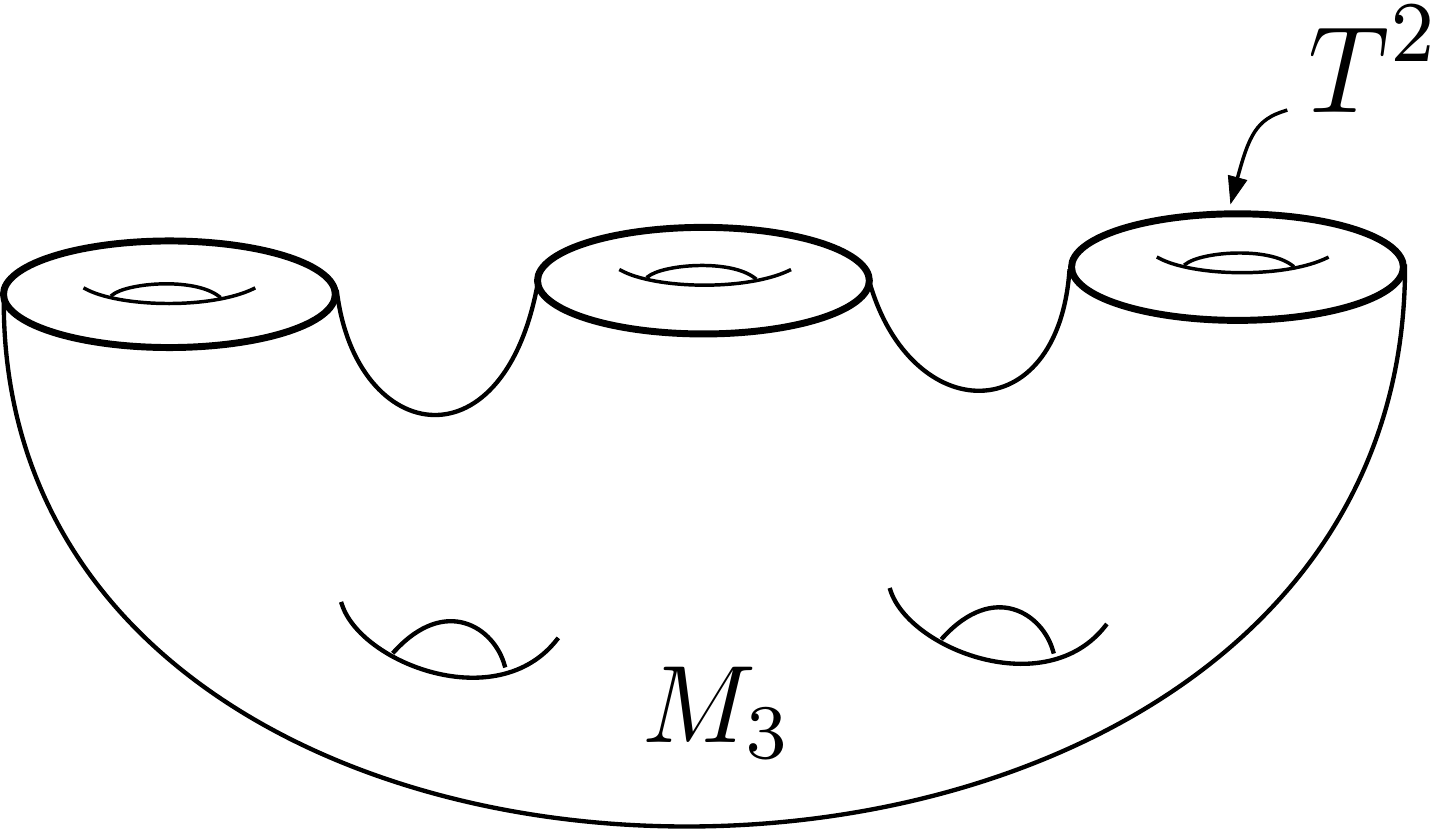}
\caption{\small{\textsf{The spatial manifold $\Sigma_n$ for $n=3$ is the disjoint union of three tori. $M_n$ is a 3-manifold such that $\pa M_n = \Sigma_n$.}}\label{fig2}}
\end{figure}

In this section, we briefly review the construction of \cite{Balasubramanian:2016sro}.  Consider Chern Simons theory with gauge group $G$ at level $k$. The action of the theory on a  3-manifold $M$ is given by
\beq
S_{CS}[A] = \frac{k}{4\pi} \int_{M} \mathrm{Tr}\,\left(A\wedge dA + \frac{2}{3} A \wedge A \wedge A\right),
\eeq
where $A= A_{\mu}dx^{\mu}$ is a gauge field (or equivalently, a connection on a principal $G$-bundle over $M$). Recall from our discussion in the previous section, that we are interested in considering disconnected spatial slices and the entanglement structure of the corresponding states. For simplicity, we consider states defined on $n$ copies of $T^2$, namely on the spatial slice  (Fig.~\ref{fig2})
\beq
\Sigma_n = \cup_{i=1}^n T^2.
\eeq
The corresponding Hilbert space is the $n$-fold tensor product $\cH^{\otimes n}$, where $\cH=\cH(T^2;G,k)$ is the Hilbert space of Chern Simons theory on a torus (for the group $G$ at level $k$).  A natural way to construct states in a quantum field theory is by performing the Euclidean path integral of the theory on a 3-manifold $M_n$ whose boundary is $\pa M_n = \Sigma_n$.  In a general field theory the state constructed in this way will depend on the detailed  geometry of $M_n$, for instance the choice of metric on $M_n$; in our situation (i.e., for a TQFT) only the topology of $M_n$ matters.  However, there are many topologically distinct Euclidean  3-manifolds with the same boundary, and the path integrals on these manifolds will construct different states on $\Sigma_n$.  Following \cite{Balasubramanian:2016sro}, we will focus on a class of such 3-manifolds called link complements, which we now briefly describe.


\begin{figure}
\centering
\includegraphics[height=5.5cm]{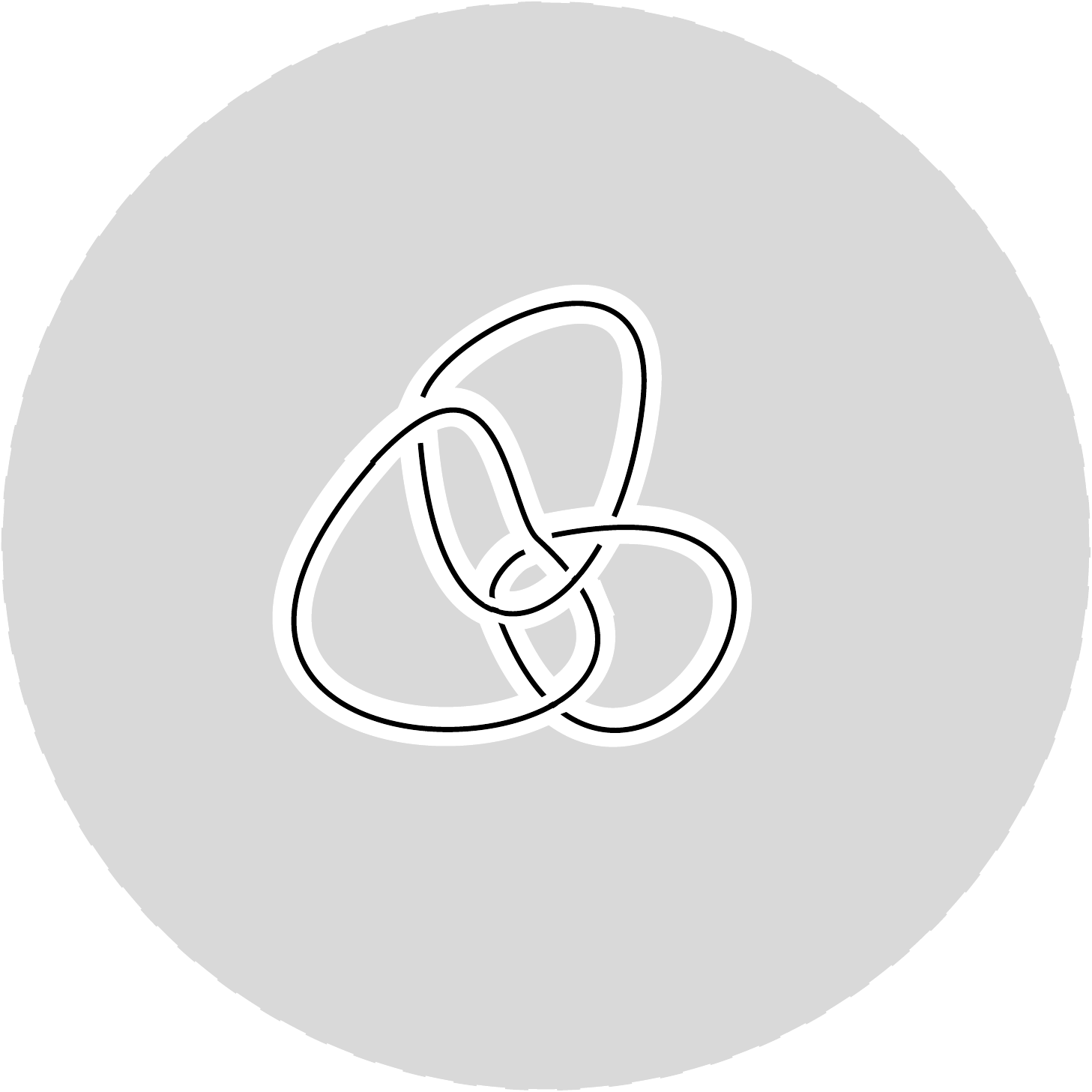}
\caption{\small{\textsf{The link complement (the shaded region) of a 3-component link (bold lines) inside the three-sphere. The white region indicates a tubular neighbourhood of the link which has been drilled out of the 3-sphere.}}\label{fig:fig3}}
\end{figure}

We start by considering an $n$-component \emph{link} in the 3-sphere $S^3$ (more generally, any connected, closed 3-manifold would do). An $n$-component link in $S^3$ is an embedding of $n$ (non-intersecting) circles in $S^3$. (Note that 1-component links are conventionally called \emph{knots}.) We will often denote a generic $n$-component link as $\cL^n$, when we do not need to choose a particular link. We will label the $n$ circles which constitute the link as $L_1, \ldots, L_n$, so $\cL^n = L_1 \cup L_2 \cup \cdots \cup L_n$. Now in order to construct the desired 3-manifold $M_n$, we remove a tubular neighbourhood $N(\cL^n)$ of the link from inside $S^3$. In other words, we take $M_n$ to be $S^3 - N(\cL^n)$, i.e., the complement of $\cL^n$ in $S^3$ (Fig.~\ref{fig:fig3}). Since $\cL^n$ is an $n$-component link, its link complement $M_n$ is a manifold with $n$ torus boundaries,\beq
\pa M_n = \cup_{i=1}^n T^2,
\eeq
which is precisely what we desired. We can therefore perform the path integral of Chern Simons theory on $M_n$, and obtain a state on $\Sigma_n$. In other words, for any given link $\cL^n$ in $S^3$, the path integral of Chern Simons theory on the link complement $M_n = S^3-N( \cL^n)$ produces a state $|\cL^n\rangle $ in the $n$-fold tensor product of the torus Hilbert space $\cH^{\otimes n}$.

The discussion above was a bit abstract, but we can give a much more concrete expression for these states in terms of a particular basis for the torus Hilbert space, which we will denote $\{ |j\rangle \}$. In order to construct the basis state $|j\rangle $, think of the torus as the boundary of a solid torus, and insert a Wilson line in the core of the solid torus along its non-contractible cycle in the representation $R_j$. For compact gauge groups, we need only consider a finite number of \emph{integrable} representations,\footnote{For instance if $G= SU(2)$, the integrable representations are are labelled by the spin $j = 0, 1/2, 1. \cdots , k/2$. } and so the Hilbert space on the torus obtained as the span of $\{|j\rangle \}$ is finite dimensional. We will not need to know further details for our present discussion, but more details can be found in \cite{Witten:1988hf}.   We can  write the state $|\cL^n\rangle $ obtained by performing the path integral of Chern Simons theory on the link complement of $\cL^n$ in terms of the above basis vectors:
\beq
|\cL^n\rangle = \sum_{j_1,\cdots, j_n} C_{\cL^n}(j_1, j_2, \cdots j_n) |j_1,j_2, \cdots, j_n\rangle,\qquad  |j_1,j_2, \cdots, j_n\rangle\equiv |j_1\rangle \otimes |j_2\rangle \otimes |j_n\rangle
\eeq
where $C_{\cL^n}(j_1,\cdots, j_n)$ are complex coefficients, which we can write explicitly as
\beq
C_{\cL^n}(j_1, j_2, \cdots j_n)= \langle j_1,j_2,\cdots j_n | \cL^n\rangle  \, .
\eeq
Operationally, this corresponds to gluing in solid tori along the boundary of the link complement $S^3-N(\cL^n)$, but with Wilson lines in the conjugate representation $R^*_{j_i}$ placed in the bulk of the $i^{th}$ torus.  Thus, the coefficients $C_{\cL^n}(j_1,\cdots j_n)$ are precisely the \emph{colored link invariants}\footnote{For the gauge group $SU(2)$, these are often called the colored Jones polynomials, after dividing by the $S^3$ partition function, which is  an overall color-independent constant. } of Chern Simons theory with the representation $R^*_{j_i}$ placed along the $i^{th}$ component of the link:
\beq
C_{\cL^n}(j_1, \cdots, j_n) = \left\langle W_{R^*_{j_1}}(L_1) \cdots W_{R^*_{j_n}}(L_n)\right\rangle_{S^3} \label{cli} \,~~ ; ~~ W_{R}(L) = \Tr_R\left(e^{\oint_L A} \right) \, ,
\eeq
where we recall that $L_i$ are the individual circles which constitute the link, namely $\cL^n = L_1 \cup \cdots \cup L_n$. Thus, the link state $|\cL^n\rangle$ encodes all the coloured link invariants corresponding to the link $\cL^n$ at level $k$. 

\begin{figure}
\centering
\includegraphics[height=5cm]{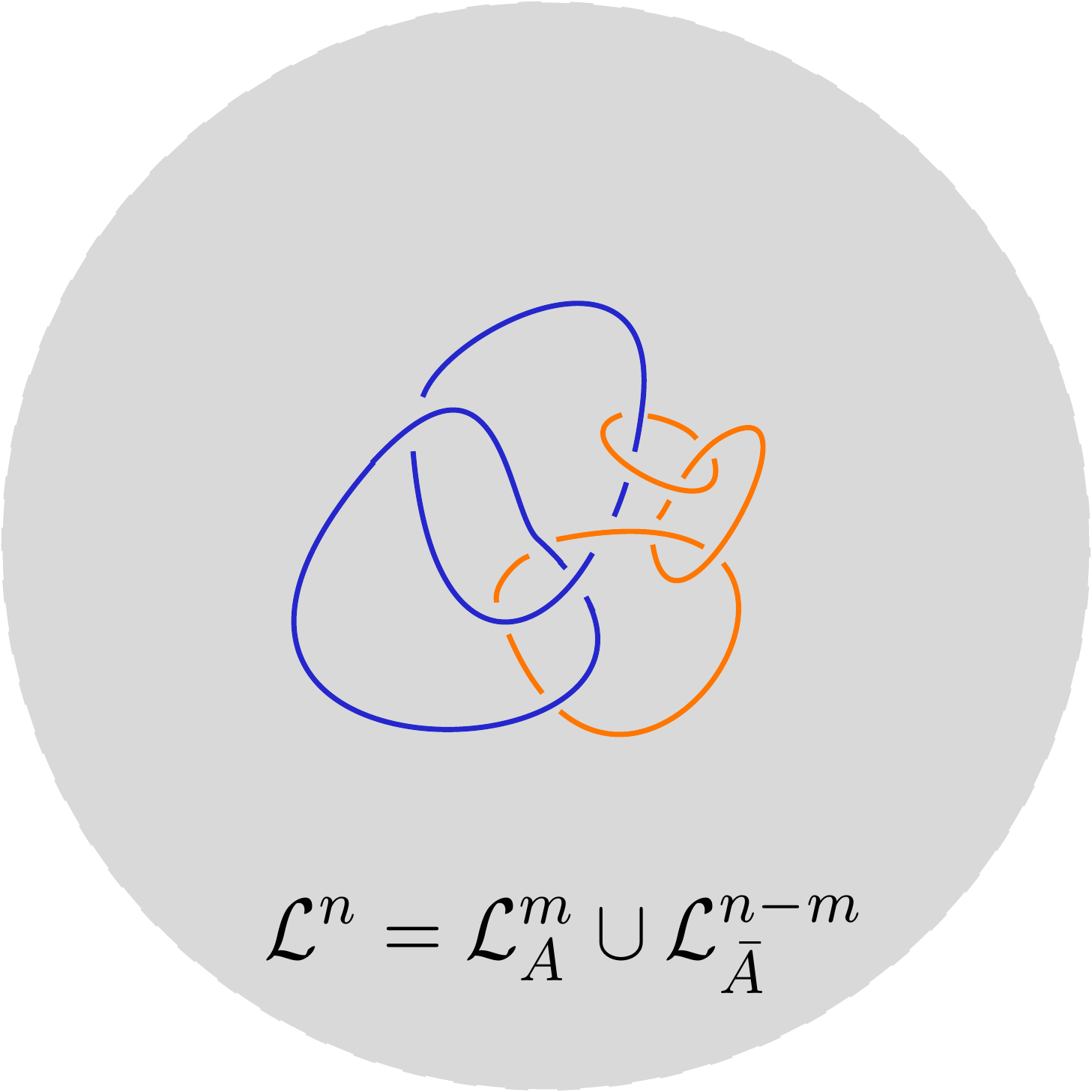} 
\caption{\small{\textsf{We can compute the entanglement between the two sublinks $\cL^m_A$ (blue) and $\cL^{n-m}_{\bar{A}}$ (orange) of $\cL^n$ by tracing out the factor corresponding to $A$ in the full state $|\cL^n\rangle$ and computing the von Neumann entropy of the resulting reduced density matrix.}}\label{fig:fig4}}
\end{figure}

The important point emphasized in \cite{Balasubramanian:2016sro} is that the above construction assigns a \emph{quantum entanglement structure}\footnote{By entanglement structure, we mean the pattern of quantum entanglement inherent in the state $|\cL^n\rangle$.} to a  link in the 3-sphere. In this paper, we will probe this entanglement structure by using standard quantum information theoretic quantities, namely entanglement entropy and separability (discussed in the previous section) upon tracing out various factors in the  state. For instance, we can compute the entanglement entropy corresponding to partitioning the $n$-component link into an $m$-component sub-link $\cL^m_A =  L_1 \cup L_2 \cup \cdots \cup L_m$ and its complement $\cL^{n-m}_{\bar{A}} = L_{m+1} \cup \cdots \cup L_n$ (see figure~\ref{fig:fig4})
\beqn
& &S_{EE}(\cL^m_{A}| \cL^{n-m}_{\bar{A}}) = -\mathrm{Tr}_{L_{m+1}, \cdots , L_n}(\rho\,\mathrm{ln}\,\rho),\nonumber\\
 & & \rho = \frac{1}{\langle \cL_n |\cL_n\rangle }\mathrm{Tr}_{L_1,\cdots, L_m}|\cL^n\rangle\langle \cL^n| ,
\eeqn
where by tracing over $L_i$ we mean tracing over the Hilbert space of the torus boundary corresponding to the circle $L_i$. Further, we can also ask about the separability properties of the reduced density matrix $\rho$ obtained by tracing out $\cL^m_A$.  We will demonstrate these ideas in the simple example of $U(1)$ Chern-Simons theory below.

Before we proceed, we point out two important facts. First, take the link $\cL^n$ to be $n$ unlinked knots. In this case, it is well-known that the coloured link-invariant in Eq.~\eqref{cli} factorizes as
\beq
 \frac{C_{\mathrm{unlink}}(j_1, \cdots, j_n)}{C_0} = \prod_{i=1}^n \frac{C_{L_i}(j_{i})}{C_0} \, ,
 \eeq
 where $C_0 = \mathcal{S}^0_0$ is the partition function of Chern-Simons theory on $S^3$. It is then clear that the state $|\cL^n\rangle$ is a product state
 \beq
 |\cL^n\rangle \propto |L_1\rangle \otimes |L_2 \rangle \otimes \cdots \otimes |L_n\rangle,
 \eeq
and hence $|\cL^n\rangle$ is completely unentangled. This suggests that the quantum entanglement of link states captures aspects of the topology  of the corresponding links. More generally, if a link \emph{splits} into two sub-links $\cL^m_A$ and $\cL^{n-m}_{\bar{A}}$, where by split we mean that there exists a 2-sphere separating one sub-link from the other, then 
\beq
|\cL^n\rangle \propto |\cL^m_{A} \rangle \otimes |\cL^{n-m}_{\bar{A}}\rangle,
\eeq
and the entanglement entropy between the two sub-links vanishes.\footnote{It is tempting to speculate that there must be a sense in which the converse statement is true as well for non-Abelian gauge groups, that is, if the entanglement entropy between two sub-links of a link vanishes, then the link splits into the two sub-links. A similar conjecture was put forth in \cite{Chun:2017hja}. } We will return to a generalized notion of separating surfaces in section \ref{mgb}, where we will use them to give an upper bound on the entanglement between sublinks.

Secondly, above, we ignored the issue of {\it framing} \cite{Witten:1988hf} of the individual knots comprising the link $\cL^n$. Intuitively, if we replace each of the circles in the link with a ribbon, then the relative linking number between the two edges of the ribbon, or \emph{self-linking}, is ambiguous. In general, to fix this ambiguity we must pick a framing for each circle, and consequently the coloured link invariants are really defined for framed links. However a different choice of framing of, let's say, the $i^{th}$ circle $L_i$ by $t$ units is equivalent to performing a $t$-fold Dehn twist on the corresponding torus.  This  corresponds to a local unitary transformation on the corresponding link state. Local unitary transformations of this type do not affect the entanglement entropies (or more general information-theoretic quantities) we are interested in. Hence, {\it  the entanglement structure is framing-independent.} 

\subsection{$U(1)$ Chern Simons Theory and Stabilizer Groups}\label{sect:AbelianCS+SG}

As a first example, let us consider $U(1)$ Chern Simons theory at level $k$. Using the expression for the link invariant from \cite{Witten:1988hf}, the wavefunction of a given $n$-component link complement state can be written as
\begin{align}
|\mathcal{L}^n\rangle = \frac{1}{k^{n/2}} \sum_{j_1, \ldots, j_n} \exp \left( \frac{2\pi i}{k} \sum_{a<b} j_a j_b \ell_{ab} \right) |j_1, \ldots, j_n\rangle, \label{eq:linkstates}
\end{align}
where the summation over the basis states $|j_1, \ldots, j_n\rangle$ is taken mod $k$ (i.e., $j_i \in \mathbb{Z}_k,\,\forall i$), and $\ell_{ab}$ is the linking number between $L_a$ and $L_b$ mod $k$. Generically, there are self-linking terms in this formula; however, they can be set to zero by a sequence of local unitary transformations (i.e., changes of framing of the links). Therefore self-linking does not affect the entanglement between sublinks, and has been omitted.

For any bipartition of the link $\mathcal{L}^n$ into two sublinks $\cL^m_{A} = L_1 \cup L_2\cdots \cup L_m$ and $\cL^{n-m}_{\bar{A}} = L_{m+1} \cup \cdots \cup L_n $, let $\bs\ell$ be the \emph{linking matrix} across the partition,
\begin{align}
\bs\ell= \begin{pmatrix} \ell_{1,m+1} & \ell_{2,m+1} & \ldots & \ell_{m,m+1} \\ \ell_{1,m+2} & \ell_{2,m+2} & \ldots &\ell_{m,m+2} \\ 
 \vdots & \vdots & \ddots & \vdots \\ \ell_{1,n} & \ell_{2,n} & \hdots & \ell_{m,n} \end{pmatrix}, \label{eq:linkmat}
\end{align}
and let $|\ker \bs\ell |$ be the cardinality of the kernel of $\bs\ell$, where the kernel is taken over the field $\mathbb{Z}_k$ of integers mod $k$. The entanglement entropy across the bipartition is given by:\footnote{For a bipartition of $\mathcal{L}^n$ into $1$-component and $(n-1)$-component sublinks, this formula reduces to
$$S_{EE} = \ln \left(\frac{k}{\text{gcd} (k, \ell_{12}, \ldots, \ell_{1n})}\right).$$}
\begin{align}
S_m = \ln \left(\frac{k^m}{|\ker \bs\ell|}\right). \label{eq:entropy}
\end{align}
This formula was derived in \cite{Balasubramanian:2016sro} by using the replica trick. A corollary of this formula is that the entanglement entropy across a bipartition vanishes if and only if the linking matrix vanishes (mod $k$). Below, we will give a different derivation of equation \eqref{eq:entropy}.

The link complement states in Abelian Chern Simons theory fall into a special class of states in quantum information theory known as \emph{stabilizer states}  \cite{Salton:2016qpp} which find important application in the theory of quantum computing and quantum error correction \cite{gottesman}.  In fact any wavefunction of the form in (\ref{eq:linkstates}) is known to be a stabilizer state \cite{gross}.  Stabilizer states have the property that they are simultaneous eigenstates of unit eigenvalue of an associated Abelian group of operators called the \emph{stabilizer group}  \cite{stabcodes1, stabcodes2, stabcodes3, linden}.  We will explicitly construct below the stabilizer group corresponding to a given link complement state.   The entanglement entropy of a sub-factor in such states is known in terms of properties of the stabilizer group \cite{linden, stabentropy}.   We will show that this formulation precisely reproduces equation (\ref{eq:entropy}) in terms of the linking matrix.

\begin{figure}[t]
\begin{center}
\includegraphics[width=0.4\textwidth]{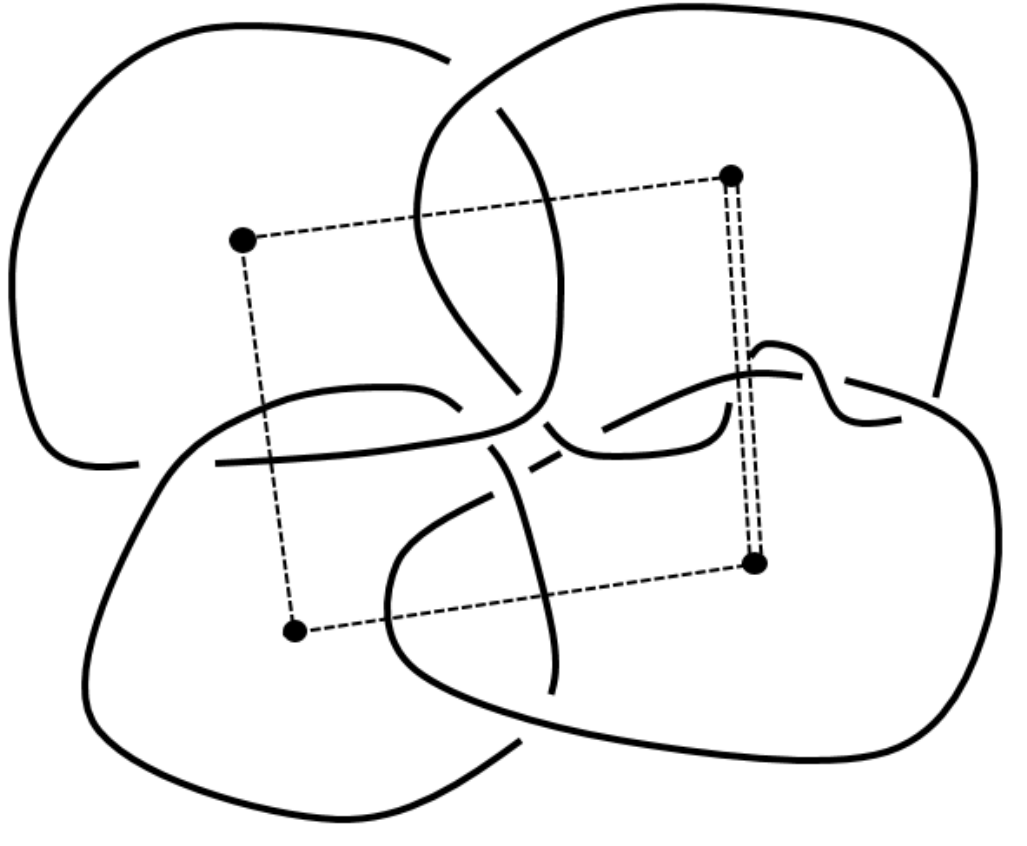}
\end{center}
\caption{\small{\textsf{A four-component link and its associated weighted graph. Each knot corresponds to one vertex in the graph. The weight of an edge (depicted here by the number of edges connecting two vertices) is the linking number between the circles corresponding to the vertices.}} \label{fig:graphlinks}}
\end{figure}

The $U(1)$ Chern-Simons states obtained from link complements in $S^3$ in fact correspond to a subclass of stabilizer states called \emph{weighted graph states} (see \cite{graphreview, graphentangle, schlingemann} for graph states on qubits, and  \cite{quditgraphs1, quditgraphs2}  for graph states on k-bits).
%
 To construct such states one starts with a weighted graph, which consists of a set of vertices $\mathcal{V}$ joined by edges $\mathcal{E}$. Each edge carries a number called a \emph{weight}; one may equivalently consider an edge of weight $w$ to correspond to $w$ edges between the same two vertices. To each vertex $a \in \mathcal{V}$, one associates the uniform superposition of states
\begin{align}
|+\rangle_a = \frac{1}{\sqrt{k}} \sum_{j \in \mathbb{Z}_k} |j\rangle_a.
\end{align}
The graph state is then built by acting with unitaries on the initial state $ |+\rangle^{\otimes n}= \otimes_{a \in \mathcal{V}} |+\rangle_a$. The unitary $U_{ab}$ creating an edge of weight $\ell_{ab}$ between vertex $a$ and vertex $b$ is specified by the following action on the basis of states for an $n$-vertex graph
\begin{align}
U_{ab} |j_1, \ldots, j_n\rangle =  \exp \left( \frac{2\pi i}{k} j_a j_b \ell_{ab} \right) |j_1 , \ldots, j_n\rangle .
\end{align}
The graph state $|\psi\rangle$ is then given by acting with the product of all unitaries corresponding to all choices of pairs of vertices. That is, for an $n$-vertex graph, the graph state is
\begin{align}
|\psi\rangle = \prod_{a,b \,\in \, \mathcal{V}} U_{ab} |+\rangle^{\otimes n}.
\end{align}
The states thus prepared are exactly the link complement states (\ref{eq:linkstates}).   One obtains the weighted graph corresponding to a given link by replacing each knot with a vertex and connecting vertices with the number of edges corresponding to the linking number between the respective knots, as in Fig.~\ref{fig:graphlinks}. The linking matrix $\bs\ell$ for any bipartition then maps to a sub-block of the adjacency matrix of the corresponding graph.

The stabilizer group of an $n$-vertex weighted graph state for arbitrary $k$ is known \cite{graphentangle} and is constructed from the discrete Heisenberg group generated by ``shift" and ``clock" operators $X$ and $Z$. In terms of the orthonormal basis $|j\rangle$ on the single-torus Hilbert space, we define $X$ and $Z$ by
\begin{align}
X|j\rangle = |j+1\rangle, \qquad Z|j\rangle = e^{\frac{2\pi i j }{k}} |j\rangle,
\end{align}
where as before $j$ is an integer mod $k$. The operators $X$ and $Z$ almost commute, except for a complex phase, $XZ = e^{-\frac{2\pi i}{k}} ZX$, and the center of the group generated by $X$ and $Z$ consists only of the $k$ complex phases $C =  \{e^{\frac{2\pi i j }{k}}, j\in \mathbb{Z}_k\}$.  The stabilizer group for weighted graph states is generated by the center of the discrete Heisenberg groups acting on the vertices of the graph (the different tori in our link complement states), and all elements of the form
\begin{align}
\left\{K_i = X_i \prod_{j\neq i} Z_j^{\ell_{ij}}  \:\biggr| \: i \in \{1\ldots n\}\right\}, \label{eq:genstab}
\end{align}
where
\begin{align}
\mathcal{O}_i = \underbrace{I \otimes I \otimes \ldots }_{i-1 \text{ operators}} \otimes \mathcal{O} \otimes \underbrace{\ldots \otimes I \otimes I}_{n-i \text{ operators}}
\end{align}
is shorthand for an operator that acts as $\mathcal{O}$ on the $i$th vertex and is otherwise the identity, so that
\begin{align}
X_i\prod_{j\neq i} Z_j^{\ell_{ij}} = \underbrace{Z^{\ell_{i1}} \otimes Z^{\ell_{i2}} \ldots}_{i-1 \text{ operators}} \otimes X \otimes \underbrace{\ldots \otimes Z^{\ell_{i(n-1)}} \otimes Z^{\ell_{in}}}_{n-i \text{ operators}}
\end{align}
is the operator which acts as $X$ on the $i$th vertex and otherwise as $Z^{\ell_{ij}}$ on the $j$th vertex.

Suppose we have a multipartite system with $n$ components such that the Hilbert space factorizes across any bipartition of the components into two sets $A$ and $\bar{A}$, $\mathcal{H} = \mathcal{H}_A \otimes \mathcal{H}_{\bar A}$. The entanglement entropy for such a bipartition of this system, for any stabilizer state $|\Psi\rangle$, can be found purely in terms of the stabilizer group $G$ of $|\Psi\rangle$. To this end, define $d_A = \prod_{x \in A} |\mathcal{H}_x|$ to be the size of the Hilbert space associated with the subset $A$, and define $G_A$ to be the set of elements in $G$ so that $G_A / C_A$ acts as the identity on $\bar A$, where $C_A = C \cap G_A$. The subgroup $G_A$ is sometimes called the \emph{local subgroup} \cite{stabentropy} as it consists of exactly the stabilizer group elements which act nontrivially only on $A$. Then the entanglement entropy is given by \cite{linden}
\begin{align}
S_A = \ln \left( \frac{d_A}{|G_A / C_A|} \right). \label{eq:stabentropy}
\end{align}

The stabilizer entropy formula (\ref{eq:stabentropy}) is very similar in appearance to the link entropy formula (\ref{eq:entropy}). For any bipartition of a link $\mathcal{L}^n$ into an $m$-component sublink $\cL^m_A$ and its complement $\cL^{n-m}_{\bar A}$, it is immediate that $d_A = k^m$, since $A$ consists of $m$ $k$-dimensional torus factors. We now show how the link complement states can be reinterpreted from the perspective of the stabilizer formalism so that $|G_A / C_A| = |\ker \bs\ell|$.

We can now explicitly compute the local subgroups of the stabilizer to obtain the entropy formula, for the general case of an arbitrary $n$-component link. Consider a general partition of the $n$ components into sets $A$ and $\bar A$. Without loss of generality we may permute the components so that $A$ consists of the first $m$ components, while $\bar A$ consists of the remaining $n-m$ components, with $m<n-m$. All elements of the stabilizer containing an $X_i$ with $i > m$ will not be in the local subgroup $G_A$, as the only way for such an element to generate elements acting trivially in the $i$th vertex is to exponentiate to the $k$th power, yielding the identity. Since the elements of $G_A$ correspond to the different ways we can multiply together generators $K_i$ of the stabilizer group to obtain the identity on $\bar A$, each unique element of $G_A$ is specified by the number of times each generator appears in a product over all generators. That is, to each element of $G_A$ we associate a set of exponents $\alpha_i$ where each $\alpha_i$ counts the multiplicity of $K_i$ in such a product. Therefore, an arbitrary element of $G_A$ can be represented as
\begin{align}
\prod_{i=1}^m \left(X_i \prod_{j\neq i} Z_j^{\ell_{ij}} \right)^{\alpha_i} = \mathcal{O}^{(m)} \otimes I^{\otimes (n-m)},
\end{align}
where $\mathcal{O}^{(m)}$ is some combination of various powers of $X$ and $Z$ operators acting on the first $m$ vertices and $I^{\otimes (n-m)}$ is the identity on $\bar A$. This is true exactly when:
\begin{align}
\prod_{i=1}^m Z^{\ell_{ij}\alpha_i} = I \label{eq:expcondition}
\end{align}
on every vertex $m +1 \leq j \leq n$. The condition (\ref{eq:expcondition}) is satisfied if and only if for each fixed $j$ the exponents vanish:
\begin{align}
\sum_{i=1}^m \ell_{ij} \alpha_i \equiv 0 \mod k.
\end{align}
The above relation on the exponents can be rewritten as the matrix system:
\begin{align}
\begin{pmatrix} \ell_{1,m+1} & \ell_{2,m+1} & \ldots & \ell_{m,m+1} \\ \ell_{1,m+2} & \ell_{2,m+2} & \ldots &\ell_{m,m+2} \\ 
 \vdots & \vdots & \ddots & \vdots \\ \ell_{1,n} & \ell_{2,n} & \hdots & \ell_{m,n} \end{pmatrix} \begin{pmatrix} \alpha_1 \\ \alpha_2 \\ \vdots \\ \alpha_m \end{pmatrix} \equiv 0 \mod k.
\end{align}
Therefore, $|G_A/C_A| = |\ker \bs\ell |$, so we find from (\ref{eq:stabentropy})
\begin{align}
S_A = \ln \left(\frac{k^m}{|\ker \bs\ell|} \right),
\end{align}
i.e., the stabilizer entropy formula is generally equivalent to the formula (\ref{eq:entropy}).  Although the linking number is a simple link invariant, the existence of closed-form formulas for the entropy as well as the stabilizer group formalism makes $U(1)$ link states a useful arena to study entanglement structures. 

\section{The Minimal Genus Separating Surface Bounds Entanglement}\label{mgb}
\newcommand{\cSS}{\Sigma}

In the previous section, we defined the notion of entanglement entropy between a sub-link and its complement inside any arbitrary link as a tool for characterizing entanglement structure of link complement states.   The first question to ask is whether  topology guarantees any general bounds on this entanglement, or vice versa.
In this section we will prove that for the gauge group $SU(2)$, the entanglement entropy across a link bi-partition gives a lower bound on the genus of surfaces in  the ambient $S^3$ which ``separate'' the two sublinks.  Reversing this, the minimal genus of surfaces separating sub-links upper bounds their entanglement entropy.  In order to explain this bound, we first define the notion of a \emph{separating surface}:

\noindent\textbf{Definition}: Given an $n$-component link $\cL^n\subset S^3$ and two sublinks $\cL^m_A$ and $\cL^{n-m}_{\bar{A}}$ such that 
$$ \cL^n = \cL^m_{A} \cup \cL^{n-m}_{\bar{A}},$$
a \emph{separating surface}  $\cSS_{A|\bar{A}} \subset S^3$ is a connected, compact, oriented two-dimensional surface-without-boundary such that: (1) $\cL^m_{A}$ in contained in the handlebody inside $\cSS_{A|\bar{A}}$,  (2) $\cL^{n-m}_{\bar{A}}$ is contained in the handlebody outside $\cSS_{A|\bar{A}}$, and (3)  $\cSS_{A|\bar{A}}$ does not intersect any of the components of $\cL^n$. 

\begin{figure}[t]
\centering
\includegraphics[height=4cm]{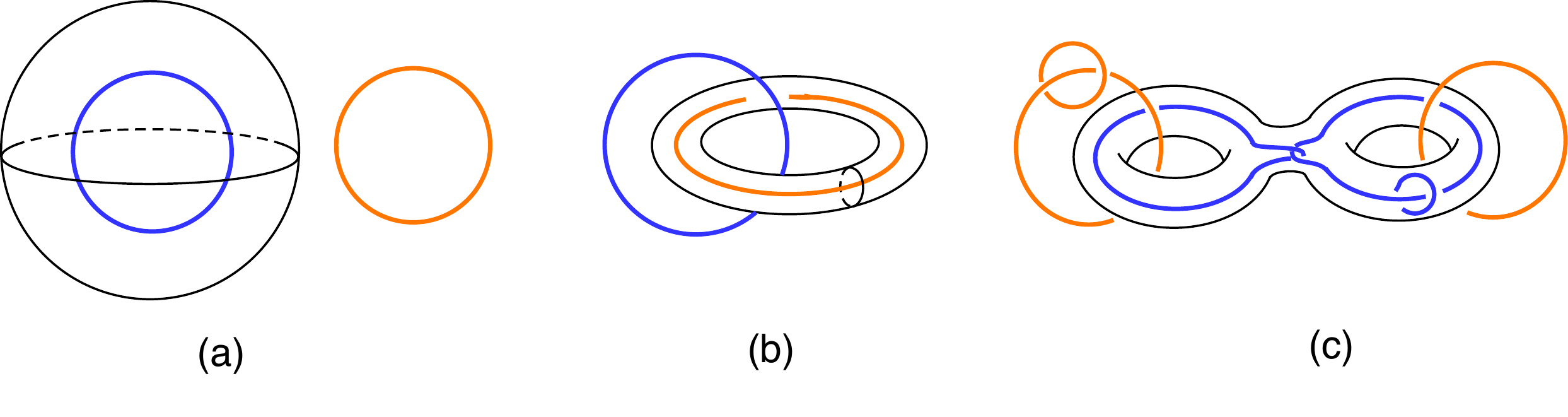}
\caption{\small{\textsf{Three examples of minimal-genus surfaces separating two subsets of links indicated in orange and blue: (a) $\mathrm{min}(g_{\cSS})=0$ for the unlink, (b) $\mathrm{min}(g_{\cSS})=1$ for the Hopf link and (c)  $ \mathrm{min}(g_{\cSS}) =2 $ for the indicated separation into two sublinks.}}\label{fig:ss}  }
\end{figure}

In other words, the separating surface gives what is a known as a Heegaard splitting of the ambient $S^3$ such that the two sublinks $\cL^m_A$ and $\cL^{n-m}_{\bar{A}}$ are separately contained in the two resulting handlebody-pieces. In order to avoid cluttering notation, we will  drop the subscripts and simply write $\cSS$ for the separating surface corresponding to a given bi-partition. The separating surface is not unique; given $ \cL^n = \cL^m_{A} \cup \cL^{n-m}_{\bar{A}},$ there are multiple topologically distinct surfaces which separate $\cL^n$ into the two sublinks. For example, in figure \ref{fig:ss}(a) we have shown the 2-sphere as a separating surface for the unlink. Of course, we could equally well draw a torus around one of the circles, and that would be an acceptable separating surface. However, there is clearly a (topologically) unique separating surface of \emph{minimal genus}; for example, the sphere is the minimal-genus separating surface for the two-component unlink. On the other hand, for the Hopf-link the minimal-genus separating surface is a torus; see figure \ref{fig:ss}(b). Similarly, fig \ref{fig:ss}(c) shows a link where the minimal-genus separating surface has genus two. Now we claim that:\vspace{0.5cm}

\noindent\textbf{Proposition 1}: \emph{Given a bi-partition $\cL^n = \cL^m_A \cup \cL^{n-m}_{\bar{A}}$, let $\mathrm{min}\,(g_\cSS)$ be the genus of the minimal-genus separating surface. Then, the entanglement entropy between $\cL_A$ and $\cL_{\bar{A}}$ provides a lower-bound on $\mathrm{min}\,(g_\cSS)$:
\beq \label{ineq}
\mathrm{min}\,(g_\cSS) \geq \frac{1}{C_{k}}\,S_{EE}(\cL_A^m |\cL^{n-m}_{\bar{A}}),
\eeq
where $C_{k} = \ln\,\left(\cS_{00}^{-2}\right)$ is a positive constant which depends on the level $k$. }
\vspace{0.5cm}

Here $\mathcal{S}_{j_1j_2} = \sqrt{\frac{2}{k+2} }\sin\left(\frac{\pi (2j_1+1)(2j_2+1)}{k+2}\right)$ is the matrix which implements the large diffeomorphism $\tau \to -\frac{1}{\tau}$ on the torus Hilbert space. We may interpret the inequality \eqref{ineq} as saying that the entanglement entropy between two sublinks gives a measure of the topological obstruction to the splitting of a link between the two sublinks. Of course, we can also flip equation \eqref{ineq} around and use it as an upper-bound on the entanglement entropy, but we will actually prove the following tighter bound below:
\beq \label{ineq2}
S_{EE}(\cL_A^m |\cL^{n-m}_{\bar{A}}) \leq \ln \left(\sum_{j=0,} ^{k/2} \frac{1}{\cS_{0j}^{2 \mathrm{min}(g_{\cSS})-2}} \right) .
\eeq
For instance in the example of the unlink, $\mathrm{min}\,(g_\cSS) = 0$, and the bound implies that the entropy is zero (which is indeed true). For the Hopf link, the bound is saturated, as the Hopf link is maximally entangled \cite{Balasubramanian:2016sro}.  There is in fact a trivial upper-bound on the entanglement entropy, namely
\beq
S_{EE}(\cL_A^m |\cL^{n-m}_{\bar{A}}) \leq \ln(k+1)\, \mathrm{min}(m, n-m) ,
\eeq
because the dimension of the Hilbert space of an m-component link is $ \left( \mathrm{dim} \cH_{T^2} \right)^m$, but the inequality \eqref{ineq2} is a non-trivial, tighter upper-bound in general, as can be checked in the example in figure \ref{fig:ss}(c). A similar bound can be derived in $U(1)$ Chern Simons theory where we have a general closed form expression, equation (\ref{eq:entropy}), for the entanglement entropy in terms of linking numbers. The bound in this case then implies\footnote{This is  of course true for an arbitrary positive integer $k$, but we can get the tightest bound by maximizing the left hand side with respect to $k$.}
\beq
m - \frac{\mathrm{ln} \,|\mathrm{ker}(\bs{\ell})|}{\mathrm{ln}\,k} \leq  \mathrm{min}\,(g_\cSS) \leq \mathrm{min}(m,n-m) .
\eeq

In order to prove the bound in equation \eqref{ineq}, we use the fact that the state corresponding to $\cL^n$ is prepared by performing the Euclidean path integral on the link complement. Now given a bi-partition of the link, let $\cSS$ be a separating surface with genus $g_\cSS$. The trick is to cut open the path integral on the link complement along $\cSS$ by inserting a complete set of states $\sum_J |J\rangle \langle J | $, where $J$ runs over a basis for the Hilbert space corresponding to $\cSS$.  Thus, the state corresponding to $\cL^n$ takes the form
\beq \label{mgb1}
|\cL^n \rangle = \sum_{j_1\cdots j_m} \sum_{j_{m+1},\cdots, j_n} \sum_J \psi_A(j_1,\cdots, j_m; J) \psi_{\bar{A}}(j_{m+1},\cdots, j_n; J) |j_1,\cdots j_m\rangle \otimes |j_{m+1},\cdots, j_n\rangle,
\eeq 
where $\psi_A$ is the path integral over the handlebody ``inside'' $\cSS$ contracted with $\langle J |$ on $\cSS$, and  $\psi_{\bar{A}}$ is the path integral over the handlebody ``outside'' $\cSS$ contracted with $| J \rangle$ on $\cSS$. We can now rewrite equation \eqref{mgb1} in the more accessible form
\beq \label{mgb2}
|\cL^n \rangle = \sum_J |\psi_A(J)\rangle \otimes |\psi_{\bar{A}}(J) \rangle,
\eeq
where the first factor is a state in the Hilbert space corresponding to $\cL^m_{A}$ and the second factor corresponding to its complement. From this expression, it is clear the reduced density matrix on $A$ takes the form
\beq \label{mgb3}
\rho_A = \sum_{J,J'} c_{J,J'}\;|\psi_A(J)\rangle \langle \psi_{A}(J') |,
\eeq
namely that it is a matrix of maximal rank equal to the dimension of the Hilbert space on $\cSS$. The dimension of the Hilbert space on a Riemann surface of genus $g_{\cSS}$ is given by \cite{VERLINDE1988360}
\beq
\mathrm{dim}\, \cH_{\cSS} = \left(\sum_{j=0,} ^{k/2} \frac{1}{\cS_{0j}^{2 g_{\cSS}-2}} \right),
\eeq
where the sum is over the integrable representations $ j =0,\frac{1}{2}, 1,\cdots, \frac{k}{2}$. The entanglement entropy is bounded by the log of the dimension of  $\rho_A$ and thus  satisfies the upper bound $S_{EE} \leq \ln\,\mathrm{dim}\,\cH_{\cSS}$. The tightest bound is obtained by choosing $\cSS$ to be the minimal-genus separating surface, in which case we obtain:
\beq 
S_{EE}(\cL_A^m |\cL^{n-m}_{\bar{A}}) \leq \ln \left(\sum_{j=0,} ^{k/2} \frac{1}{\cS_{0j}^{2 \mathrm{min}(g_{\cSS})-2}} \right) .
\eeq
For $\mathrm{min}(g_{\cSS}) \geq 1$, we can obtain a simpler inequality by noting that $\cS_{0j} \geq \cS_{00}$ for all $j$, so we can make the replacement $\cS_{0j} \to \cS_{00}$ in each term above. Using $\ln(k+1) \leq \ln\left(\cS_{00}^{-2}\right)$ to further simplify, we finally obtain the advertised result:
\beq 
S_{EE}(\cL_A^m |\cL^{n-m}_{\bar{A}}) \leq \mathrm{min}(g_{\cSS}) \ln \left(\cS_{00}^{-2} \right) .
\eeq

The minimal-genus bound we have proven above is similar in spirit to the Ryu-Takayangi formula for the entanglement entropy of a subregion in a holographic conformal field theory. In that case one is instructed to find a minimal area surface which hangs into the AdS-bulk and is homologous to the CFT subregion, while in the present case we are instructed to minimize the genus of a surface which separates the two sublinks. However, the Ryu-Takayanagi formula is an equality (as opposed to a bound); in this sense, our bound is more closely analogous to the minimal-area bound on the entropy of subregions in the MERA tensor network construction of states in conformal field theory \cite{Nozaki:2012zj, Pastawski:2015qua}. In our case, we arrived at the minimal-genus bound by cutting open the Euclidean path integral along the minimal-genus separating surface, while the minimal-area bound in MERA is proved by cutting open the tensor network along the minimal-area cut through the network. This suggests that our path integral arguments might have natural generalizations to more non-trivial quantum field theories (i.e., beyond topological theories), although we expect the argument would have to deal with the standard ultraviolet divergences of quantum field theory as soon as we move away from the TQFT limit.

\section{Entanglement Structure of Torus and Hyperbolic Links } \label{sec4}
In the previous section we demonstrated a general topological bound on the entanglement entropy between sublinks.   This bound shows that if the sublinks can be split, i.e., separated by a 2-sphere, then they must have vanishing entanglement.   In this section we consider {\it non-split links} for which there is no bipartition separated by a 2-sphere.  Such links can have inherently multi-partite entanglement, because there is no sublink that must disentangle from the remainder.   Here, inspired by the two classes of intrinsically 3-qubit entanglement patterns (GHZ and W, see Introduction), we will focus on a limited issue, i.e., whether partial traces over some link components produce a separable state on the remainder.   This leads to the following definition:
%
%

%

\noindent\textbf{Definition}: {A state with three or more sub-factors will be said to have \emph{GHZ-like} entanglement if the reduced density matrix obtained by tracing out any sub-factor is mixed (i.e., has a non-trivial von Neumann entropy) but is separable on all the remaining sub-factors. A state with three or more sub-factors will be said to have \emph{W-like} entanglement if the reduced density matrix obtained by tracing out any sub-factor is mixed but not always separable on the remaining sub-factors.}

Two important topological classes of non-split links are the \emph{torus links} (i.e., links which can be drawn on the surface of a torus) and the \emph{hyperbolic links} (i.e., links whose link complement supports a hyperbolic structure). In fact, every non-split, alternating, prime link is either a torus link or a hyperbolic link \cite{MENASCO198437}.\footnote{Here ``alternating" means that crossings along any circle alternate above and below, and ``prime" means that that the link is not a connected sum of other links.}    We will study the entanglement structure in these two classes of links. 

\subsection{Torus links}
\begin{figure}[t]
\centering
\includegraphics[height=3.5cm]{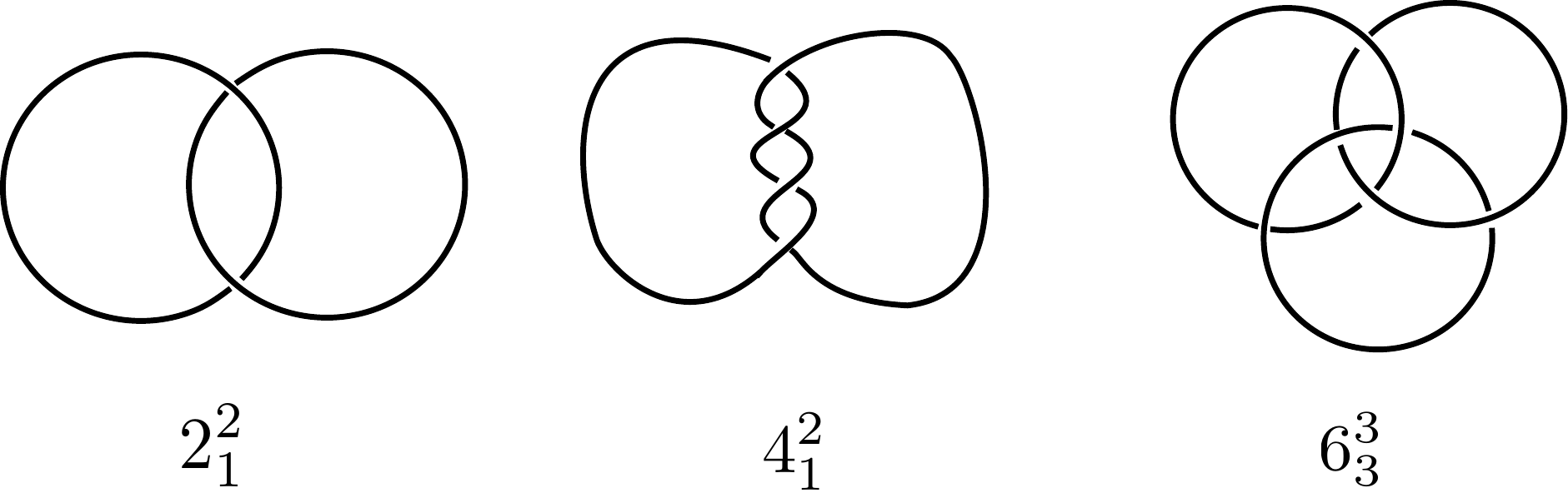}
\caption{\small{\textsf{ Some examples of torus links labeled using Rolfsen notation. \label{fig:tl} }}}
\end{figure}

Torus links, namely links which can be embedded on the surface of a two dimensional torus (without self intersection), are  an important topological class.   Some examples include $2^2_1$ (the Hopf link), $4^2_1$, and $6^3_3$  (see Fig.~\ref{fig:tl}). In fact the entanglement structures of these examples were already studied in \cite{Balasubramanian:2016sro}, where it was shown that in $SU(2)$ Chern-Simons theory the Hopf link is maximally entangled and the three-component link $6^3_3$ is GHZ-like. We will prove the following  general result:

\noindent\textbf{Proposition 2}: \emph{All torus links with three or more components have a GHZ-like entanglement structure.}

The proof will show that the state corresponding to any torus link always takes the form
\beq \label{tl0b}
|\cL^n \rangle= \sum_{\ell} \lambda_{\ell} (\cL^n) \widetilde{|\ell\rangle }\otimes \widetilde{|\ell\rangle }\otimes \cdots \widetilde{|\ell\rangle },
\eeq
where $\{|\widetilde{\ell}\rangle \}$ is a particular basis for the torus Hilbert space to be defined below (compare with Eq.~\ref{GHZstate} for the GHZ state on three qubits).   It is clear from  \eqref{tl0b} that tracing out any sublink leaves us with a separable density matrix on the remainder. This result establishes a direct connection between a topological property of links and a quantum information-theoretic property of the corresponding states.  We now give a short proof of Proposition 2. 

\begin{figure}[h]
\centering
\includegraphics[width=.2\textwidth]{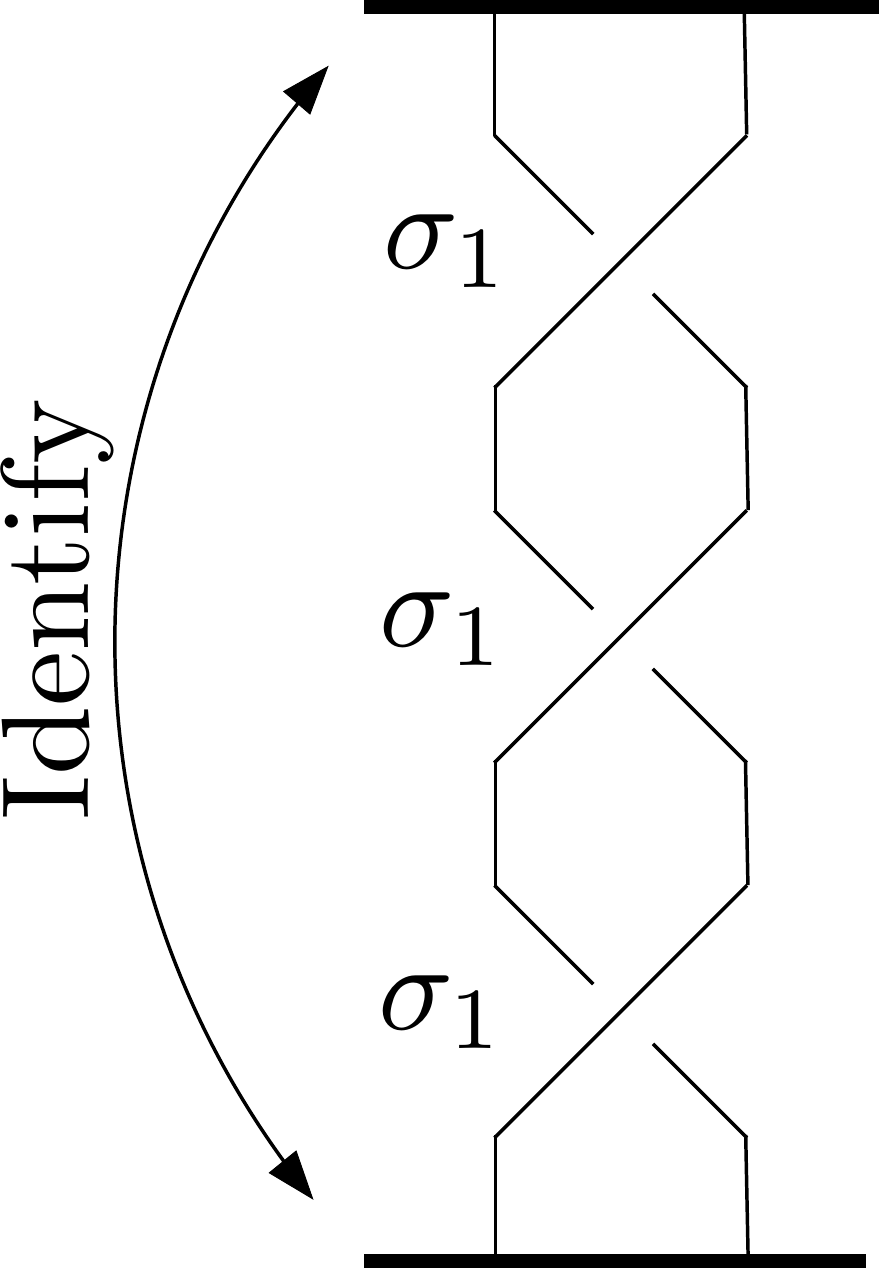}\includegraphics[width=.3\textwidth]{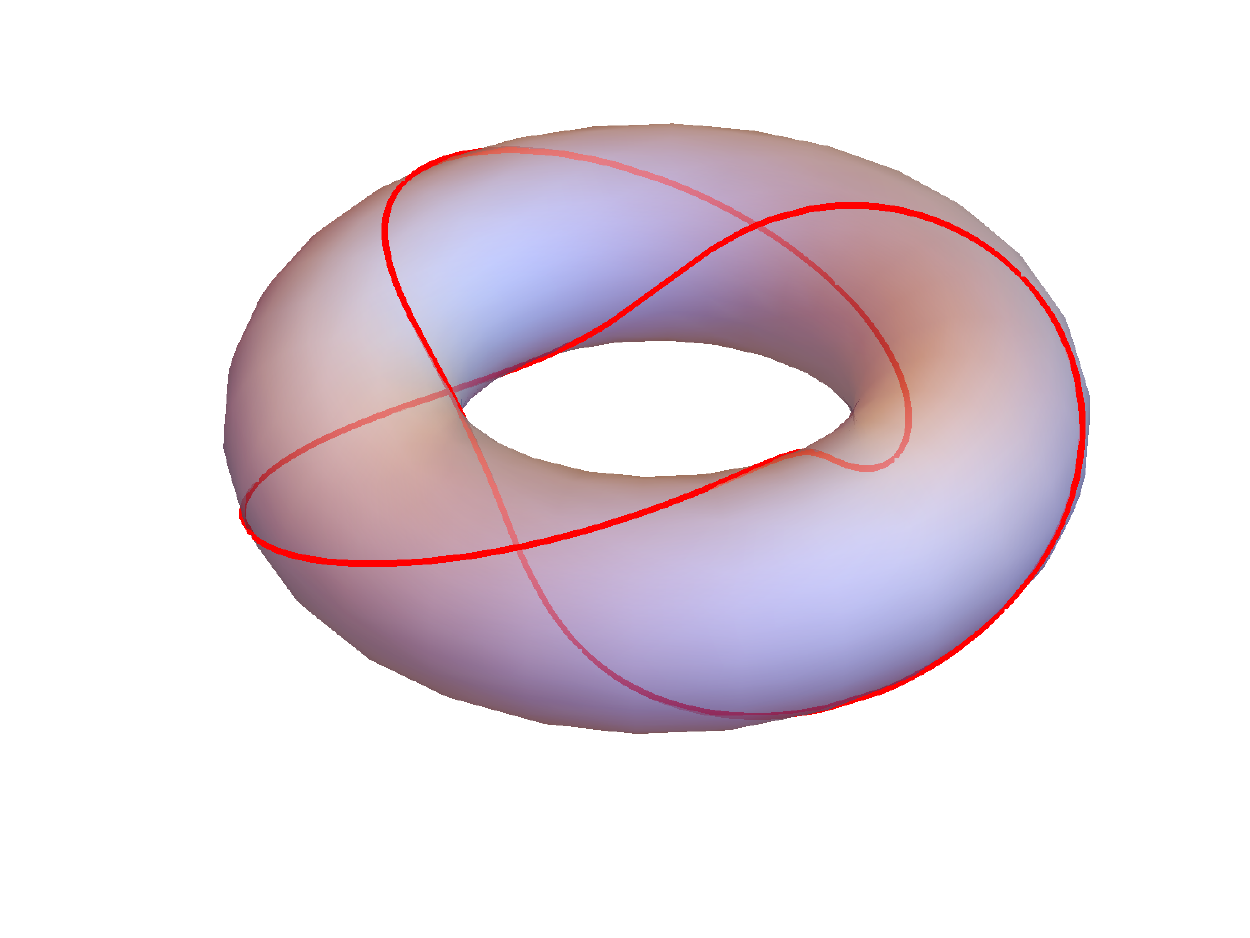}\includegraphics[width=.2\textwidth]{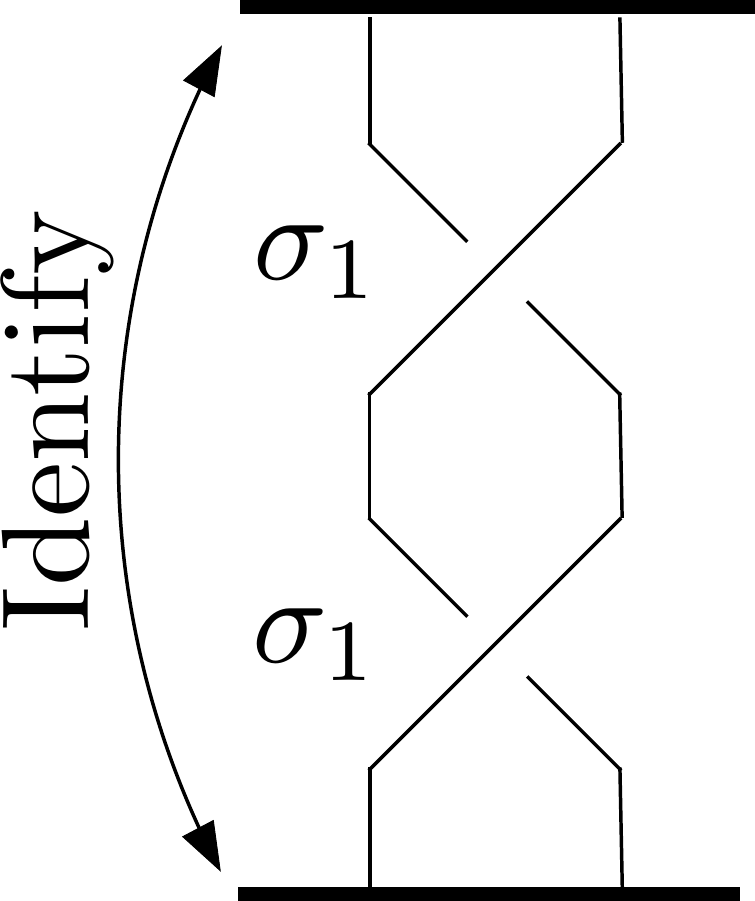}\includegraphics[width=.3\textwidth]{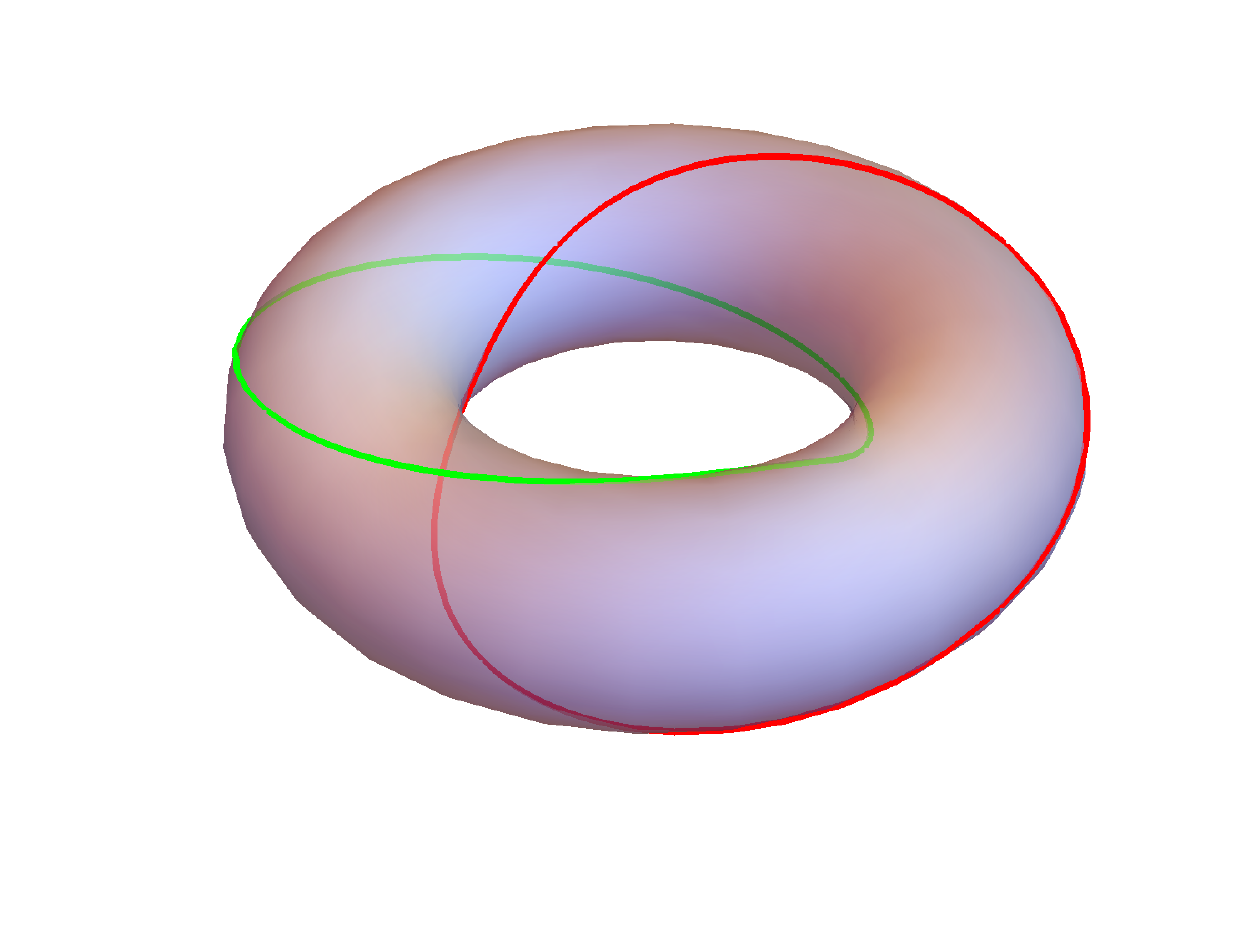}
\caption{\small{\textsf{(Left) The trefoil knot as a (2,3) torus knot braid and drawn on the surface of a torus.  (Right) The Hopf link as a (2,2) torus link braid and drawn on the surface of a torus.}}}
\label{fig:tlbraid}
\end{figure}

Torus links are characterized by two integers $P$ and $Q$. Given two integers $(P,Q)$, the $(P,Q)$ torus link (often referred to as $T(P,Q)$) can be constructed as the closure of the braid $(\sigma_1\sigma_2\ldots\sigma_{P-1})^Q$ acting on $P$ strands. Here, $\sigma_i$ denotes the crossing of strand $i$ over $i+1$.  This is illustrated in Fig. \ref{fig:tlbraid} for $P=2$.  We may take $0 < P < Q$ without loss of generality. It is easy to see that when $P$ and $Q$ are relatively prime, the closure of the braid results  in a 1-compnent link (a knot) which wraps around the torus longitude of the torus $P$ times, and around the meridian $Q$ times.  However, when $\text{gcd}(P,Q)=n$ the closure of the braid will result in an $n$ component link, each component of which wraps around the torus longitude and meridian $P/n$ and $(Q/n)$ times respectively.

Let us warm up by examining torus links in $U(1)$ Chern-Simons theory.  This is a useful exercise since, as described in section \ref{sect:AbelianCS+SG}, we possess an exact closed-form formula for the link state of generic $U(1)$ link that depends only on the mutual linking numbers.  In fact, for a $(P,Q)$ torus link, examination of the braid word closure shows that the mutual linking numbers are \emph{homogeneous}: i.e., $\ell_{ab}=\ell,\;\;\forall a\neq b$ (for a particular choice of orientation of the individual knots).  A counting of the crossings\footnote{That is, let $\mathcal C$ be the total number of crossings excluding self crossings: $\mathcal C=2\sum_{i<j}\ell=\ell\,n(n-1)$. In the braid word $(\sigma_1\sigma_2\ldots\sigma_{P-1})$ there are $P-1$ crossings, $P/n-1$ of which are self crossings.  Repeating the braid word $Q$ times yields $\mathcal C=Q(P-1-P/n+1)$.  Equating the two gives the stated result.} in the braid diagram reveals $\ell=\frac{PQ}{n^2}$.  As such the Abelian link state for a $(P,Q)$ torus link is given by
\begin{equation}
|\mathcal L(P,Q)\rangle=\frac{1}{k^{n/2}}\sum_{j_1,\ldots,j_n}\exp\left(\frac{2\pi i}{k}\ell\sum_{a<b}j_a\,j_b\right)|j_1,\ldots, j_n\rangle \label{eq:toruscomplete}
\end{equation}
Up to phase, $e^{\frac{\pi i}{k}\ell\sum_a j_a^2}$, acting on each tensor factor (which can be removed by a local unitary) this state can be written as
\begin{equation}
|\mathcal L(P,Q)\rangle=\frac{1}{k^{n/2}}\sum_{j_1,\ldots,j_n}\exp\left(\frac{\pi i}{k}\ell\left(\sum_{a}j_a\right)^2\right)|j_1,\ldots, j_n\rangle 
\end{equation}
Let us denote the total charge (mod $k$) of the basis element $|j_1,\ldots, j_n\rangle$ by $\hat j=\sum_{a=1}^n j_a$.  We can rewrite $|\mathcal L(P,Q)\rangle$ in terms of $\hat j$ by imposing a periodic delta function:
\begin{equation}
|\mathcal L(P,Q)\rangle=\frac{1}{k^{n/2+1}}\sum_{q=1}^k\sum_{\hat j=1}^k\sum_{j_1,\ldots,j_n}\exp\left(\frac{\pi i}{k}\ell\hat j^2\right)\exp\left(\frac{2\pi i}{k}q\left(\hat j-\sum_{a}j_a\right)\right)|j_1,\ldots, j_n\rangle
\end{equation}
(The sum on $q$ imposes the delta function.) We now see that $j_a$-dependent coefficients can be removed by the local unitary change of basis
$|q\rangle = \frac{1}{\sqrt{k}}\sum_{j} \exp(-{2\pi i \over k} q j) | j \rangle$. 
  The state is then unitarily equivalent to 
\begin{equation}
|\mathcal L(P,Q)\rangle=\frac{1}{k}\sum_{q=1}^k\sum_{\hat j=1}^k\exp\left(\frac{2\pi i}{k}q\,\hat j+\frac{\pi i\ell}{k}\hat j^2\right)|q,q,\ldots, q\rangle\equiv \sum_{q=1}^k\lambda_q(P,Q)|q,q,\ldots, q\rangle.
\end{equation}
proving (\ref{tl0b}). Thus we see that torus links in $U(1)$ Chern-Simons are GHZ-like. An alternate proof of this result can also be given using the fact that the wavefunction (\ref{eq:toruscomplete}) describes a \emph{complete} graph state where all edges have weight $\ell$ \cite{graphentangle, graphreview, PhysRevA.95.052340}.\footnote{The proof works by showing that the GHZ state is unitarily equivalent to the state corresponding to the \emph{star graph} by a sequence of discrete Fourier transforms (Hadamard transforms, when $k=2$). Then, a unitary graph operation called \emph{local complementation} takes the star graph to the complete graph and vice versa. } 

We now move on to $SU(2)$ Chern Simons theory. In particular, given an  n-link $\cL^n \subset S^3$, the corresponding state (in the canonical basis introduced previously) is given by
\begin{equation} \label{tl0}
|\cL^n \rangle = C_0 \sum_{j_1\ldots j_n}J_{j_1,\cdots j_n}(\cL^n) |j_1\ldots j_n\rangle.
\end{equation}
where for $SU(2)$, the colors $j_i$ run over $0, \frac{1}{2}, 1, \cdots, \frac{k}{2}$, $C_0$ is an overall constant (more precisely it is the $S^3$ partition function) and the wavefunction $J_{j_1,\cdots j_n}(\cL^n)$ is the colored Jones polynomial.  Proceeding generally, we note that a systematic way to evaluate the colored Jones polynomials of torus links   is to take a $(P,Q)$ $n$-component link with representations $j_1,\ldots, j_n$ and to fuse them sequentially using the Chern-Simons fusion rules into a $(P/n,Q/n)$ torus knot summed over representations with the appropriate fusion coefficients \cite{Isidro:1992fz, Labastida:2000yw, Brini:2011wi}.\footnote{This fusion is possible because all the components of torus links are simply braiding along one of the cycles of the defining torus.}   
We refer the reader to the above references for further details, and merely state here the result for the colored Jones polynomial:\footnote{We are omitting an overall phase proportional to the central charge.  Additionally \cite{Brini:2011wi} writes the final link invariant in terms of the quantum dimension which differs from (\ref{tl1}) by a factor of $\mathcal S^0_0$, a matter of normalization.}
\beq \label{tl01}
J_{j_1, \cdots, j_n} (P,Q) = \sum_{\ell_1, \ell_2,\cdots} N_{j_1j_2\ell_1}N_{\ell_1j_3\ell_2} \cdots N_{\ell_{n-2}j_n\ell_{n-1}} J_{\ell_{n-1}}(P/n, Q/n).
\eeq
where $N_{ijk}$ are the fusion coefficients. Further using the Verlinde formula \cite{VERLINDE1988360}
\beq
N_{ijk} = \sum_{\ell} \frac{\cS_{i\ell} \cS_{j\ell} \cS_{k\ell}}{\cS_{0\ell}},
\eeq
where, as before, $\mathcal{S}_{j_1j_2} = \sqrt{\frac{2}{k+2} }\sin\left(\frac{\pi (2j_1+1)(2j_2+1)}{k+2}\right)$ is the \emph{unitary} matrix which implements the large diffeomorphism $\tau \to -\frac{1}{\tau}$ on the torus Hilbert space, we can rewrite the colored Jones polynomial in the form
\beq\label{tl1}
J_{j_1\ldots j_n} (P,Q)=\sum_{\ell}\sum_{j_s}\frac{1}{\left(\mathcal S_{0\ell}\right)^{n-1}}\mathcal S_{\ell j_1}\mathcal S_{\ell j_2}\ldots \mathcal S_{\ell j_n} \mathcal S_{\ell j_s} J_{j_s}(P/n,Q/n).
\eeq
Here
\beq
J_{j_s}(P/n, Q/n) =\sum_{j_p}C^{j_p}_{j_s}(P/n)\,\mathcal S_{0j_p}\, e^{i2\pi \frac{Q}{P} h_p}
\eeq
is the colored Jones polynomial for the $(P/n, Q/n)$-torus knot, $h_p$ is the conformal primary weight of the representation $j_p$ and the coefficients $C_{j_s}^{j_p}$ are defined as 
\begin{equation}
\mathrm{Tr}_{j_s}\left(\hat U^m\right)=\sum_{j_p}C^{j_p}_{j_s}(m)\, \mathrm{Tr}_{j_p}\left(\hat U\right),
\end{equation}
for any holonomy $\hat{U}$. (For instance, $C^{j_p}_{j_s}(1)=\delta^{j_p}_{j_s}$.) For our purposes, these details are not too important; what is important however is the structure of the colored Jones polynomial in equation \eqref{tl1}, which we can rewrite as
\beq\label{tl2}
J_{j_1\ldots j_n} (P,Q)=\sum_{\ell} \frac{1}{\left(\mathcal S_{0\ell}\right)^{n-1}}\mathcal S_{\ell j_1}\mathcal S_{\ell j_2}\ldots \mathcal S_{\ell j_n}  f_{\ell}(P,Q)
\eeq
where 
\beq
f_{\ell}(P,Q) = \sum_{j_s}\mathcal S_{\ell j_s} J_{j_s}(P/n,Q/n).
\eeq
Using equations \eqref{tl0} and \eqref{tl2}, we then find that the state corresponding to a generic $(P,Q)$-torus link takes the form
\beqn \label{tl3}
|\cL_{(P,Q)}\rangle  &=& C_0\sum_{\ell} \frac{1}{\left(\mathcal S_{0\ell}\right)^{n-1}}f_{\ell}(P,Q)\, |\widetilde{\ell}\rangle \otimes |\widetilde{\ell} \rangle \otimes \cdots |\widetilde{\ell}\rangle \nonumber\\
&\equiv & \sum_{\ell} \lambda_{\ell}(P,Q)\, |\widetilde{\ell}\rangle \otimes |\widetilde{\ell} \rangle \otimes \cdots |\widetilde{\ell}\rangle
\eeqn
where we have defined the new basis $|\widetilde{j} \rangle = \sum_{j'} \cS_{jj'} |j'\rangle$, which is related to the old basis by a local unitary transformation ($\cS\cdot \cS^{\dagger} = \cS^{\dagger}\cdot \cS = 1$). This is convenient because we are interested here in understanding the entanglement structure, which remains invariant under such a local (i.e., acting on each local tensor factor) change of basis. We have thus arrived at our desired result, equation \eqref{tl0b}.

Now let us investigate what happens when we trace over some subset of links.  Since it is obvious from \eqref{tl3} that the state is invariant under permutations of the ordering of the components, without loss of generality we can trace over the final $n-r$ links, leaving a reduced density matrix on the remaining $r$ links.  It is easy to see that in doing so the reduced density matrix remains diagonal. The normalized reduced density matrix for any subset of $r$ links can be written as
\begin{equation} \label{tlrdm}
\hat\rho_{r|n-r}(P,Q)=\sum_{\ell}\Lambda_{\ell}(P,Q)|\widetilde{\ell},\cdots, \widetilde{\ell}\rangle\langle \widetilde{\ell},\cdots, \widetilde{\ell}|
\end{equation}
with the normalized eigenvalues
\beq
\Lambda_l(P,Q)=\frac{\left|\lambda_l(P,Q)\right|^2}{\sum_l \left|\lambda_l(P,Q)\right|^2}
\eeq
This is a completely separable density matrix on the remaining sub-links indicating that the entanglement in the full link had a GHZ-like structure.
Note that the eigenvalues, $\Lambda_l(P,Q)$ encode the specifics of the underlying torus link.  However these eigenvalues \emph{are independent of how many factors have been traced out,} as long as $0<r<n$.  Therefore the multi-boundary entanglement entropy for torus links takes the particularly simple form
\begin{equation}
 S_{r|n-r}(P,Q)=-\sum_l\Lambda_l(P,Q)\log\Lambda_l(P,Q)
\end{equation}
for all $0<r<n$. In addition, it is clear that the reduced density matrix \eqref{tlrdm} is separable for any choice of bi-partition. In other words, the reduced density matrix does not contain  \emph{any quantum entanglement}; all the quantum entanglement in the original state was genuinely multi-partite and GHZ in character.

While the arguments above were presented in the case of the gauge group $SU(2)$, we expect these arguments to generalize to arbitrary compact gauge groups. This is because the crux of the derivation (equations \eqref{tl01}, \eqref{tl1} and \eqref{tl2}) merely used the fusion rules for Chern-Simons theory (i.e., the Verlinde formula) together with the unitarity of $\cS$. Since these are general properties of Chern-Simons theory with compact gauge groups, our arguments will be valid for general compact groups. This concludes our derivation of the result that the entanglement structure of all torus links is GHZ-like. 

\subsection{Hyperbolic links}
\begin{figure}
\centering
\includegraphics[height=4cm]{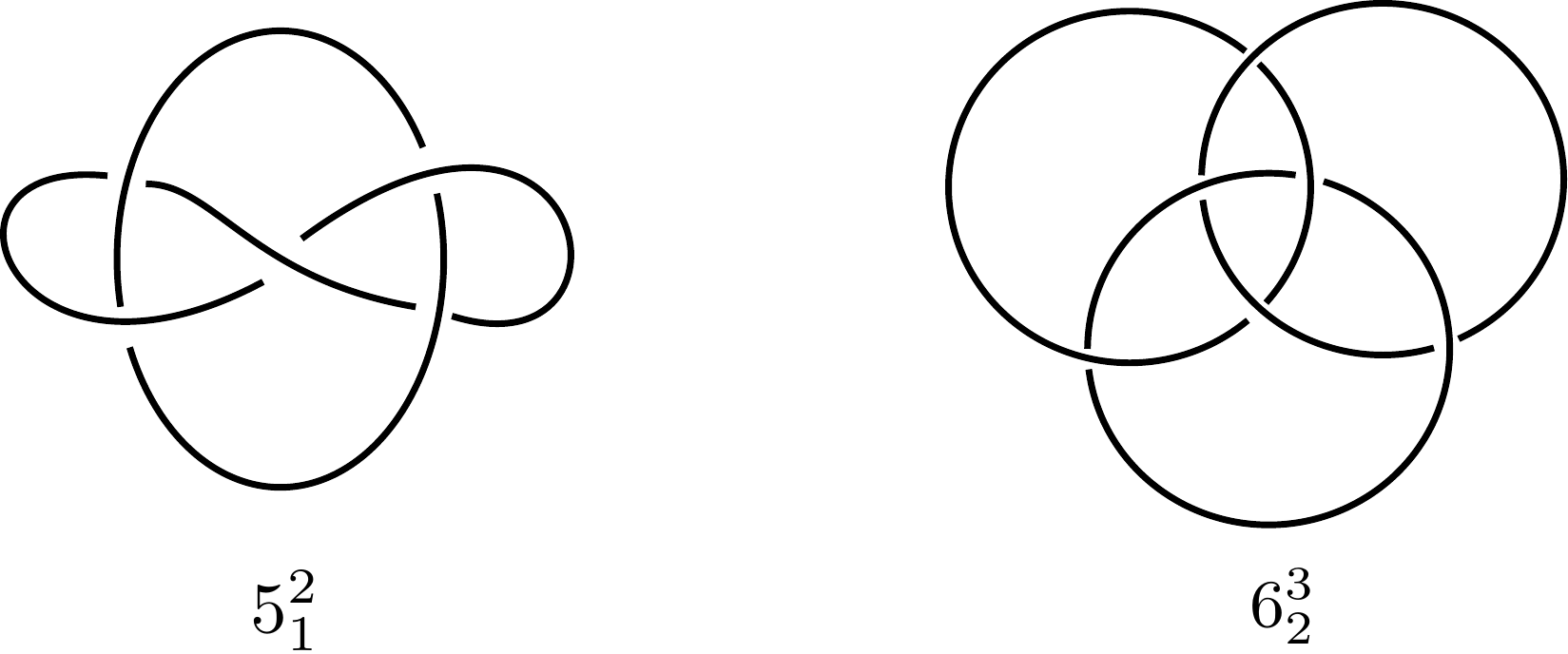}
\caption{\small{\textsf{Two examples of hyperbolic links: Whitehead link (left) and Borromean rings (right).}}\label{fig:HL}}
\end{figure}
Next we consider \emph{hyperbolic links}, whose link complements admit a complete hyperbolic structure, namely a geodesically complete metric with constant negative curvature. Some examples of hyperbolic links, the Whitehead link and the Borromean rings (Fig.~\ref{fig:HL}), were already studied in the $SU(2)$ theory in \cite{Balasubramanian:2016sro}.  It was shown there that the Borromean rings have a W-like entanglement structure. 
(The Whitehead link has only two components and thus does not have multi-party entanglement.)
In this section, we will present further evidence suggesting that hyperbolic links are generically W-like. 

In order to proceed, on the knot theory side we need to compute the colored Jones polynomials of hyperbolic links. Unfortunately, to the best of our knowledge, there is not much known about the general structure of these polynomials for hyperbolic links (as compared to torus links for instance), so we proceeded case-by-case by looking at several three-component hyperbolic links.  Our strategy was to compute the colored Jones polynomials by writing the link in terms of a braid representation.  We then used the monodromy properties of chiral conformal blocks in $SU(2)_k$ Wess-Zumino-Witten theory. This method was explained in detail in \cite{Kaul:1993hb} and reviewed in the appendix A of \cite{Balasubramanian:2016sro}, so we will not repeat the details here. Actually, we found it convenient to use a slight variant of this technique, where we first expressed the link as a braid in $S^2\times S^1$ (with an extra circle which does not braid with the original link), and then used surgery to obtain the colored Jones polynomial in $S^3$ (as explained in \cite{Witten:1988hf}).\footnote{This procedure was numerically implemented using \emph{Mathematica}.} 

On the quantum information theory side, we need an efficient way to detect whether the reduced density matrix obtained after tracing out one of the factors is separable. A useful information theoretic quantity along these lines is the \emph{entanglement negativity} \cite{PhysRevLett.77.1413,PhysRevA.65.032314, Rangamani:2015qwa}. For a given (possibly mixed) density matrix $\rho$ on a bi-partite system, let us start by defining the partial transpose $\rho^{\Gamma}$:
\beq
\langle j_1,j_2 | \rho^{\Gamma}  | \tilde{j}_1, \tilde{j}_2 \rangle = \langle \tilde{j}_1,j_2 | \rho  | j_1, \tilde{j}_2 \rangle.
\eeq
which also satisfies $\mathrm{Tr}(\rho^\Gamma) = 1$ just like $\rho$.  If $\rho^{\Gamma}$ has any negative eigenvalues, then this necessarily implies that the density matrix $\rho$ is not separable \cite{PhysRevLett.77.1413}. The sum of the negative eigenvalues can be captured by the {\it entanglement negativity} $\mathcal{N}$, which is defined as
\beq
\mathcal{N}  =\frac{|| \rho^{\Gamma} || - 1}{2},
\eeq
where $|| A ||  = \mathrm{Tr}\left(\sqrt{A^{\dagger}A} \right)$ is the trace norm. A non-zero value of $\mathcal{N}$ therefore necessarily implies that the reduced density matrix is non-separable. In our context, the results in the previous section (Proposition 2) together with the fact that all alternating, prime, non-split links are either torus or hyperbolic \cite{MENASCO198437}, imply the following corollary: 

\noindent \textbf{Corollary 3}: \emph{If a prime, alternating, non-split link has entanglement negativity $\mathcal{N} > 0$ for some bipartion of some proper sublink,\footnote{A proper sublink of $\cL$ is a sublink which is not equal to $\cL$.} then the link is hyperbolic.}

This provides a quantum information theoretic sufficient-but-not-necessary condition for a link to be hyperbolic. Importantly, the negativity can be computed directly from the colored Jones polynomial. In table \ref{tab1} we present entanglement negativities for twenty three 3-component non-split links in SU(2) Chern-Simons theory, eighteen of which are hyperbolic (i.e., have non-zero hyperbolic volumes).  More precisely, we traced out one of the tensor factors in the link, and then computed the entanglement negativity of the reduced density matrix on the remaining two factors. We see that all the hyperbolic links in the table have a non-zero entanglement negativity, showing that the corresponding reduced density matrices are \emph{not} separable. Therefore, these links have a W-like entanglement structure. Furthermore, all the non-hyperbolic links in table \ref{tab1} have zero negativity, which is (at the very least) consistent with our discussion in the previous section. The results presented in table \ref{tab1} suggest the conjecture that hyperbolic links in Chern-Simons theories with a compact non-Abelian gauge group for generic\footnote{It can happen that at special values of $k$, certain hyperbolic links degenerate to a product structure. This happens for instance at $k=1$ for the Borromean rings, but for $k\geq 2$ the Borromean rings are W-like. We will encounter another example of this in $SL(2,\bC)$ Chern Simons theory in the limit $G_N \to 0$.} values of the level $k$ always have a W-like entanglement structure. It would be interesting to prove this statement.

\begin{table}[t]
\centering 
\begin{tabular}{| l |c | r|}
\hline
 \textbf{Link Name} & \textbf{Negativity} $\mathcal{N}$ at $k=3$ & \textbf{Hyperbolic volume} \\
 \hline
 L6a4  & 0.18547 & 7.32772 \\
 L6n1 & 0 & 0 \\
 L8a16 & 0.097683 & 9.802 \\
 L8a18 & 0.189744 & 6.55174 \\
 L8a19 & 0.158937 & 10.667 \\
 L8n3 & 0 & 0 \\
 L8n4 & 0.11423 & 5.33349 \\
 L8n5 & 0.18547 & 7.32772 \\
 L10a138 & 0.097683 & 10.4486 \\
 L10a140 & 0.0758142 & 12.2763 \\
 L10a145 & 0.11423 & 6.92738 \\
 L10a148 & 0.119345 & 11.8852 \\
 L10a156 & 0.0911946 & 15.8637 \\
 L10a161 & 0.0354207 & 7.94058 \\
 L10a162 & 0.0913699 & 13.464 \\
 L10a163 & 0.0150735 & 15.5509 \\
 L10n77 & 0 & 0 \\
 L10n78 & 0.189744 & 6.55174 \\
 L10n79 & 0.097683 & 9.802 \\
 L10n81 & 0.15947 & 10.667 \\
 L10n92 & 0.11423 & 6.35459 \\
 L10n93 & 0 & 0 \\
 L10n94 & 0 & 0 \\
 \hline
\end{tabular}
\caption{\small{\textsf{Negativity in $SU(2)$ Chern Simons at level $k=3$ for various three-component links alongside the hyperbolic volume of the complement manifold. The hyperbolic volumes were computed using the SnapPy program \cite{SnapPy} (where zero volume implies that the given link is not hyperbolic). The colored Jones polynomials were computed using braiding representations for these links together with monodromy properties of conformal blocks in the $SU(2)$ WZW theory. In order to compute the negativity, we first trace over one of the tensor factors, and then compute the negativity of the reduced density matrix on the remaining two factors. 
}}
\label{tab1}}
\end{table}

\section{Hyperbolic Links in $SL(2,\bC)$ Chern Simons Theory}  \label{sec5}
More can be done with hyperbolic links if we complexify the gauge group to $SL(2,\bC)$.
In this case, in a certain asymptotic limit we can use the known behavior of the colored Jones polynomial of a hyperbolic link in terms of the hyperbolic geometry of its link complement. In this section, we present some results in this direction. 

We begin with a brief review of $SL(2,\bC)$ Chern Simons theory (see \cite{Witten:1989ip, Gukov:2003na,Dimofte:2009yn, Witten:2010cx, Dimofte:2016pua} for detailed expositions on the subject). The fundamental field in the theory is the gauge field $\cA$ which takes values in the Lie algebra $\mathfrak{sl}(2,\bC)$. The path integral for the $SL(2,\bC)$ Chern Simons theory is given by 
\beq
Z = \int D\cA \, D\bar{\cA} \;e^{iS[\cA, \bar{\cA}]},
\eeq
\beq \label{act0}
S= \frac{t}{8\pi} \int \mathrm{Tr}\left(\cA\wedge d\cA + \frac{2}{3}\cA\wedge \cA \wedge \cA\right)+\frac{\bar{t}}{8\pi} \int \mathrm{Tr}\left(\bar\cA\wedge d\bar\cA + \frac{2}{3}\bar\cA\wedge \bar\cA \wedge \bar\cA\right),
\eeq
where $\bar{\cA}$ is the complex conjugate of $\cA$. If we write $t = k+is$ and $\bar{t} = k - is$, then $k$ must be an integer, and $s$ has to be either purely real or purely imaginary, results which follow from unitarity \cite{Witten:1989ip, Gukov:2003na}. The case $s\in \mathbb{R}$ corresponds to gravity in Lorentzian signature with a positive cosmological constant, while $s = -i\sigma,\,\sigma \in \mathbb{R}$ corresponds to Euclidean gravity with a negative cosmological constant. We are interested here in this latter case. To be a bit more explicit, we pick $SU(2)$ as a real form of $SL(2,\bC)$, and write $\cA = \omega + \frac{i}{\ell}e$, where both $\omega$ and $e$ are $\mathfrak{su}(2)$-valued connections. It is natural to interpret $\omega$ as the \emph{spin-connection} and $e$ as the \emph{vielbein} of general relativity. Then the action \eqref{act0} becomes (setting $\ell =1 $ for simplicity) 
\beqn
S &=& \frac{k}{4\pi} \int \mathrm{Tr}\left(\omega\wedge  d\omega + \frac{2}{3}\omega\wedge \omega\wedge \omega- e\wedge de - 2\omega \wedge e\wedge e \right)\nonumber\\
&-& \frac{s}{2\pi}\int \mathrm{Tr}\left(e\wedge  d\omega + e\wedge \omega\wedge \omega  - \frac{1}{3} e\wedge e\wedge e\right),
\eeqn
up to a total derivative term. Since the integrand of the path integral is $e^{iS}$, if we are interested in Euclidean signature we must take $s = -i\sigma$ with $\sigma \in \mathbb{R}$. In this case, the exponent in the path integral is of the form 
$$\exp\left( - \frac{\sigma}{4\pi} \int \sqrt{g} \left(-R +2\Lambda \right)+\frac{ik}{4\pi}I_{grav\,CS}\right),$$
where the first term above is precisely the Einstein-Hilbert action with negative cosmological constant, while the second term is the gravitational Chern Simons term. We can then regard $\sigma$ as being proportional to the inverse of the Newton constant, $\sigma = \frac{1}{4G_N}$. In this paper, we will be interested in the asymptotic limit $\sigma \to \infty$. For simplicity, we will also set $k=0$. 

An important aspect of Chern-Simons theories with  non-compact gauge groups  such as $SL(2,\bC)$ is that the Hilbert space on $T^2$ is infinite-dimensional (see discussion below).   In the case of compact gauge groups the multi-boundary entanglement was finite for two reasons: (1) the Hilbert space on $T^2$ is finite dimensional, and (2) the multi-boundary entanglement does not involve spatial  cuts across which the entanglement can diverge.  In the case of $SL(2,\bC)$ the second property still holds, so the only potential source of divergence is the infinite size of the Hilbert space.  However, as we will see below, at least for hyperbolic links and in the asymptotic limit $\sigma \to \infty$, the multi-boundary entanglement in $SL(2,\bC)$ Chern-Simons remains finite because of the Gaussian structure of the wavefunctions.


\subsection*{Multi-Boundary States}
 Let us consider an $n$-component hyperbolic link $\cL^n$ inside $S^3$. As before, the link complement $S^3- N(\cL^n)$ is a 3-manifold with $n$ torus boundaries. The path integral of $SL(2,\bC)$ Chern-Simons theory on the link complement then produces a state in the $n$-fold tensor product of the torus Hilbert space, which as before we label $|\cL^n\rangle$. In order to proceed, we need a basis for the torus Hilbert space in the $SL(2,\bC)$ theory. Following \cite{Gukov:2003na} let us denote (the conjugation classes of) the holonomies of $\cA$ around the meridian and longitude of the torus by $\rho(\gamma_m)$ and $\rho(\gamma_\ell)$ respectively.   (The holonomies will play the role of the Wilson lines that provided a nice basis for the torus Hilbert space when the gauge group was compact.)   It is possible to write $\rho(\gamma_m) $ and $\rho(\gamma_\ell)$ in the form 
$$\rho(\gamma_m)=\left(\begin{matrix} m & \star \\ 0 & m^{-1}\end{matrix}\right),\;\;\;\; \rho(\gamma_{\ell})=\left(\begin{matrix} \ell & \star \\ 0 & \ell^{-1}\end{matrix}\right),$$ 
where $m,\ell \in \bC^*$ and $\star$ is one if $m = \ell$, and zero otherwise. Let us also introduce the notation $m = e^u$ and $\ell = e^v$ for convenience. Classically, $m$ takes values in $\bC^*$ (namely the complex plane minus the origin), so $\mathrm{Re}\,u \in \mathbb{R}$ while $\mathrm{Im}\,u $ is $2\pi$-periodic (i.e., $u$ coordinatizes a cylinder); the same holds for $\ell$ and $v$. Together $(m,\ell)$ or equivalently $(u,v)$ parametrize the classical phase space.\footnote{Typically, one also quotients by the Weyl group, but following \cite{Gukov:2003na} we will suppress this quotient.} Clearly, the phase space is non-compact, indicating that the Hilbert space upon quantization will be infinite-dimensional. At $ k =0$ and $\sigma \to \infty$, we can choose a polarization such that wavefunctions are $L^2$ functions of $u$, and independent of $v$ (i.e., in quantum mechanics we take the wavefunctions to be functions of half of the phase space coordinates, in this case $u$). In other words, the Hilbert space is spanned by the basis $\{ |u \rangle\}$, with $e^u \in \bC^*$ as in the classical case above, with the standard norm $\langle u | u'\rangle = \delta^{(2)}(u-u')$. Consequently, a basis for the $n$-fold tensor product of the torus Hilbert spaces takes the form $|u_1,\cdots, u_n\rangle = |u_1\rangle \otimes |u_2\rangle \otimes \cdots |u_n\rangle$. 

We can now write the state $|\cL^n\rangle$ as 
\beq \label{sp0}
 |\cL^n\rangle = \int d^2u_1 \cdots \int d^2u_n \langle u_1, \cdots, u_n |\cL^n \rangle |u_1,\cdots, u_n\rangle,
 \eeq
where the integration regions are over cylinders as explained above. The wavefunction $\langle u_1, \cdots, u_n |\cL^n \rangle$ is given by the path integral of Chern Simons theory on the link complement $S^3 - N(\cL^n)$, with boundary conditions which fix the boundary meridional holonomies to be $m_i$'s. In the $\sigma \to \infty$ limit, we can use the saddle point approximation to the path integral to write
\beq
\langle u_1, \cdots, u_n |\cL^n \rangle = \sum_{\a} e^{-\frac{\sigma}{\pi} V^{(\a)}(u_1,\cdots, u_n)+ \cdots }
\eeq
 where $\a$ labels the various saddle points which contribute to the path integral in the $\sigma \to \infty$ limit. These naturally correspond to locally hyperbolic ``geometries'' on $S^3 - N(\cL^n)$ (loosely speaking, solutions to Einstein's equations with negative cosmological constant, but more precisely flat $SL(2,\bC)$ connections). The function $V^{(\a)}$ is the corresponding oriented \emph{volume} of the link complement, while $\cdots$ denote higher quantum invariants which will not be relevant for us in this work. While it is not easy to write down the metrics explicitly, these geometries can nevertheless be constructed fairly explicitly by gluing together \emph{ideal tetrahedra} in hyperbolic space, following the seminal work of Thurston \cite{Thurston} (see also \cite{purcell2017hyperbolic, ratcliffe2006foundations, Dimofte:2009yn}).  Details of this construction  and an explicitly worked example are given in Appendix A.

On a general branch $\a$, the geometry associated to the flat connection labelled by the holonomies $(u_1,\cdots, u_n)$  is not geodesically complete \cite{Thurston}. However, there always exists one branch, often called the \emph{geometric branch} denoted by $\a = \mathrm{geom}$, which at the point $u_i=0 \, \forall i$ gives rise to a complete hyperbolic structure.\footnote{Recall that hyperbolic links are defined by the existence of at least one complete hyperbolic structure on the link complement.} In fact, by the Mostow rigidity theorem, such a complete hyperbolic structure is unique. The corresponding volume $V^{(\mathrm{geom})}(0)$ is therefore a topological invariant. This invariant famously appears in a certain asymptotic (double-scaling) limit of the colored Jones polynomial, a statement which goes by the name of the \emph{volume conjecture} \cite{Kashaev:1996kc, 1999math......5075M, Gukov:2003na, Dimofte:2010ep}. Away from $u_i = 0$, the hyperbolic structure on the link complement (at a generic point $u_i$) is not complete; it is nevertheless a legitimate $SL(2,\bC)$ flat connection that we must sum over in the path integral. 

\begin{figure}[t]
\centering
\includegraphics[height=4.5cm]{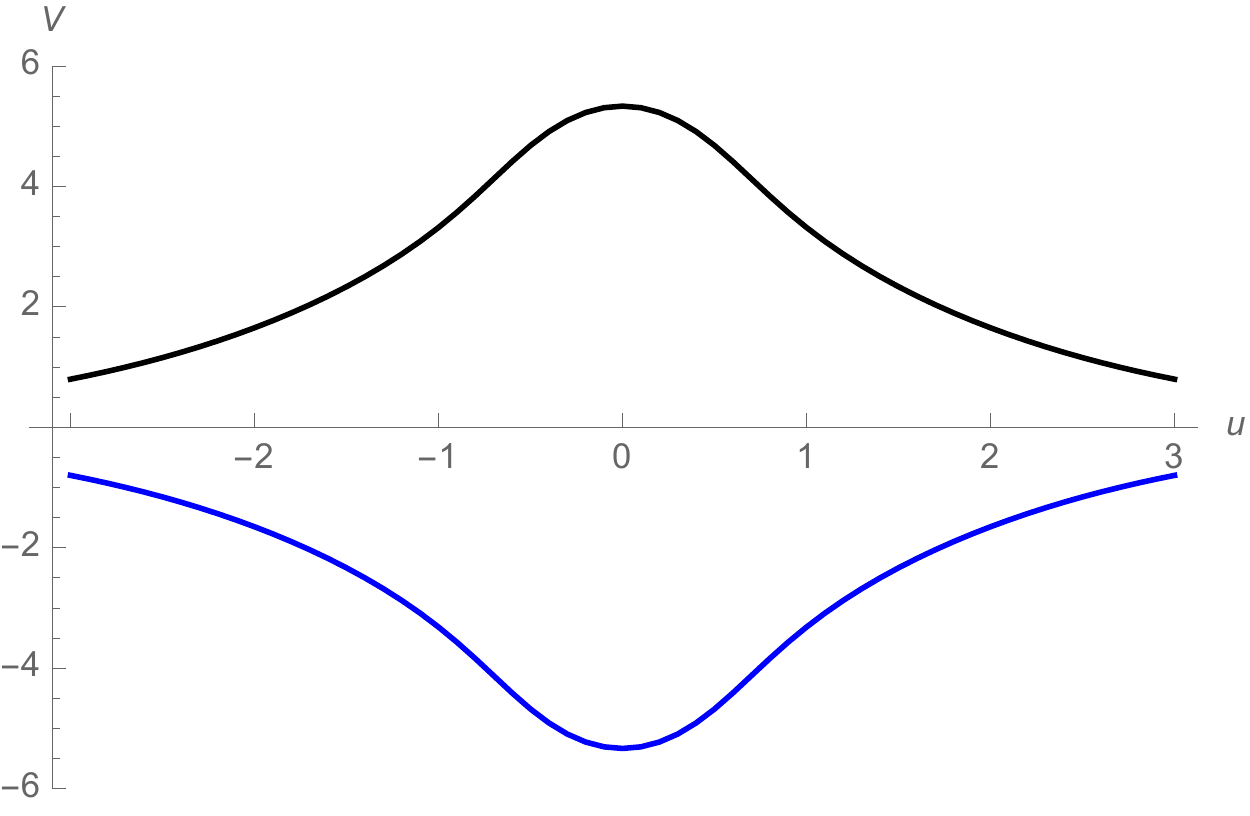}
\caption{\small{\textsf{The volume function on the geometric (black) and conjugate (blue) branches for a two component link L6a1. The coordinate $u$ here is the real part of one of the parameters on moduli space.} \label{l6a1} }}
\end{figure}

For our purposes however, a different branch will be relevant. Note that in the $\sigma \to \infty$ limit, the dominant contribution in \eqref{sp1} comes from the branch with the most negative volume (see Fig. \ref{l6a1}).\footnote{Recall that these volumes are oriented and thus can have either sign, as explained in Appendix A.} In other words, the branch most relevant for our purposes in the one which contains the global minimum of the volume function $V^{(\a)}(u_i)$, if one exists. There is indeed one such branch, which turns out to be the \emph{conjugate} of the geometric branch $\a = \overline{\mathrm{geom}}$ \cite{1999InMat.136..623D}, which then dominates the sum over saddle points.  (Appendix A explains the sense in which this branch is ``conjugate'' to the geometric one.)  On this branch the volume is minimized (most negative) at $u_i = 0$.   Then from equation \eqref{sp0}, we find that in the $\sigma \to \infty $ limit,
\beq \label{sp2}
 |\cL^n\rangle \sim \mathcal{C} \int d^2u_1 \cdots \int d^2u_n e^{-\frac{\sigma}{\pi} V^{(\overline{\mathrm{geom}})}(u_1,\cdots,u_n)} |u_1,\cdots, u_n\rangle,
 \eeq
where $\mathcal{C}$ is the normalization constant, and we use the $\sim$ symbol to indicate that we are only focussing on the conjugate-geometric branch; we will drop the superscript $\overline{\mathrm{geom}}$ from now on to prevent cluttering notation. Exploiting the $\sigma \to \infty $ limit further, we can expand the volume function around $u_i = u_i^* + \frac{1}{\sqrt{\sigma}} \delta u_i$, where $u_i^*=0$ is the location of the global minimum of the volume function  and $\delta u_i \in \bC$. Since we are expanding around $u_i=0$, we may as well drop the $\delta$s (with the understanding that now the $u_i$s are general complex numbers) and write
\beq
 V(u_1,\cdots,u_n) = V(0) + \frac{1}{2\sigma} H_{ij; ab} u^a_i  u^b_j + \cdots.
\eeq
where $a,b$ run over the real and imaginary parts of $ u_i$. This expansion was first studied in the seminal work of Neumann and Zagier \cite{NEUMANN1985307}; we now briefly review some of their results. The expansion is conveniently formulated in terms of a holomorphic function $\Phi(u_i)$ called the \emph{Neumann-Zagier potential}. Importantly, $\Phi$ is an even function of all of the $u_i$'s, and therefore takes the form
\beq
\Phi(u_i) = \sum_i \frac{\tau^{(0)}_i}{\sigma} u_i^2 +\frac{1}{2\sigma^2} \sum_{i,j} A_{ij}u_i^2 u_j^2 + \cdots
\eeq
where $\tau^{(0)}_i$ is the modular parameter of the $i$th torus boundary metric induced from the complete hyperbolic structure at $u_i=0$. In terms of the Neumann-Zagier potential, we can write the volume of the link complement as
\beq
V(u_i) = V_0 -\frac{1}{4} \sum_i \mathrm{Im} \left(u_i\overline{v_i} \right) +\frac{1}{8}\sum_{k=0}^{\infty} (k-2) \mathrm{Im}\left( \Phi_{(k)} (u_i) \right),
\eeq
where 
\beq
v_i = \frac{1}{2}\frac{\pa \Phi}{\pa u_i},
\eeq
and $\Phi_{(k)}$ is the degree $k$ part of $\Phi$. Therefore, the volume function takes the form
\beq\label{eq:Vexpand}
V(u_i) = V_0 +\frac{1}{4\sigma}\sum_i \mathrm{Im}\left(\tau_i^{(0)}\right) u_i\bar{u_i}- \frac{1}{4\sigma^2} \sum_{i,j} \mathrm{Im}\left(u_i\overline{A_{ij}u_iu_j^2}-\frac{1}{2} A_{ij}u_i^2u_j^2\right) + \cdots.
\eeq
The state \eqref{sp2} then takes the form
\beq \label{sp3}
 |\cL^n\rangle \sim \frac{\mathcal{C}e^{-\frac\sigma\pi V_0} }{\sigma^n} \int d^2 u_1 \cdots \int d^2 u_n e^{-\frac{1}{4\pi}\sum_i \mathrm{Im}\left(\tau_i^{(0)}\right) u_i\bar{u_i}+ \frac{1}{4\pi\sigma} \sum_{i,j} \mathrm{Im}\left(u_i\overline{A_{ij}u_iu_j^2}-\frac{1}{2} A_{ij}u_i^2u_j^2\right) + \cdots} |\frac{1}{\sqrt{\sigma}}u_1,\cdots, \frac{1}{\sqrt{\sigma}} u_n\rangle,
 \eeq
 where the normalization $\mathcal{C}$ can be systematically determined in terms of $\sigma, \tau_i^{(0)}$ etc. Note that at leading order in $\sigma$, the wavefunction we have obtained is a Gaussian wavepacket centered at the global minimum. Importantly, the quadratic part of the exponential is diagonal in the various torus boundaries. This is a direct consequence of the fact that the Neumann-Zagier potential is an even function of the $u_i$'s. Thus, we conclude:
 
\noindent\textbf{Proposition 4}: \emph{In the limit $\sigma \to \infty$, the state corresponding to any hyperbolic link in $SL(2,\mathbb{C})$ Chern Simons theory is a completely product state, i.e., the entanglement entropy for any sub-link vanishes.}

However, this is really a somewhat trivial manifestation of the fact that the volume is an even function of the $u_i$s. In order to study the entanglement structure, we must then back off from the $\sigma \to \infty$ limit and look at the $1/\sigma$ terms in the exponential.  These indeed introduce entanglement between the various torus boundaries. \emph{The off-diagonal elements of the matrix $A_{ij}$ therefore control the entanglement structure of the state at leading order\footnote{Note that the leading order correction to the entropy appears at order $\frac{1}{\sigma^2}$; the same is true of the entanglement negativity. Another subtlety to keep in mind while computing such corrections is that away from $\sigma = \infty$, some of the moduli might take on discrete values.} in $\frac{1}{\sigma}$}  or equivalently at leading order in the Newton constant $G_N$. The reader might worry that since we are expanding the volume to $O(1/\sigma)$, we must also include \emph{quantum corrections} to the path integral at this order. This is indeed correct; however, the quantum corrections are themselves even functions of $u_i$ \cite{Dimofte:2009yn}, and therefore at the order we are working only shift the \emph{diagonal} quadratic terms 
 $$\sum_i \mathrm{Im}\,(\tau_i^{(0)})u_i\bar{u}_i \to \sum_i \mathrm{Im}\,(\tau_i^{(0)} ) u_i\bar{u}_i+ \frac{1}{\sigma} \sum_i \left(\alpha u_i u_i + \beta u_i \bar{u}_i + \gamma \bar{u}_i\bar{u}_i\right),$$ 
This shift in the quadratic part is diagonal in the torus boundaries, and therefore does not introduce any entanglement. Therefore, we may safely focus on the matrix $A_{ij}$ coming from the Neumann-Zagier potential.  This matrix is computable, case-by-case, from SnapPy data.  In Appendix A, we perform this calculation for the Borromean rings (L6a4) and find
\begin{equation}\label{eq:ABorr}
A_{ij}^{Borr.}=i\,64\left(\begin{array}{ccc}-1/3&1&1\\1&-1/3&1\\1&1&-1/3\end{array}\right).
\end{equation}
The off-diagonal components indicate that at quartic order this link state is not a product state of each component.

Unfortunately, beyond doing this link-by-link, this is as far as we can go for now; apart from examples of explicit computation (see \cite{aaber2010closed} for one such example), to our knowledge there has been no systematic study of the matrix $A_{ij}$ in the mathematics literature.  An interesting question is whether it is possible to show in generality (from the properties of $A_{ij}$) that hyperbolic links have a W-like entanglement structure. We leave this for future work. We end here with a couple of remarks: first, it is important to note that while the detailed computation uses specific geometric structures on the link complement, the entanglement entropy is a \emph{topological invariant} (by construction)! This is exactly analogous to the fact that the hyperbolic volume of the link complement is a topological invariant -- the explanation lies in the Mostow-Prasad rigidity theorem about the uniqueness of the complete hyperbolic structure. Second, we have seen above that the entanglement structure in the $\sigma \to \infty$ limit is essentially controlled by the matrix $A_{ij}$. This is very reminiscent of Abelian Chern Simons theory, where the entanglement structure is controlled entirely by the linking matrix. Indeed, the $\sigma \to \infty$ limit is in some sense a classical limit, albeit a subtle one.\footnote{For instance, it is well known that taking the $k\to \infty$ limit (while keeping the colors fixed) of colored link invariants in non-Abelian Chern Simons theory reduces these colored link invariants to the Abelian ones (which are only sensitive to linking numbers). However, if one takes the double scaling limit $j\to \infty,\; k \to \infty$ with $2j/k$ fixed, then the asymptotic behaviour is very different. Note that the entanglement entropy is indeed sensitive to such a double-scaling limit.} Nevertheless, we have discovered that in this limit, a new matrix appears to control the entanglement structure. 

\section{Discussion}
In this paper, we have presented various results on the information theoretic properties of the colored Jones polynomial of multi-component links. We first reviewed the simple case of $U(1)$ Chern Simons theory, where we recast and clarified previous results from \cite{Balasubramanian:2016sro} in terms of the theory of stabilizer groups. Then we presented several new results for non-Abelian Chern-Simons theory: (i) We proved that the entanglement entropy between two sublinks of an arbitrary link provides a lower bound on the minimum genus Heegaard splitting which separates the two sublinks, and thus gives a measure of the topological obstruction for a link to be split, (ii) We then studied the entanglement structures of two topological classes of links, namely torus and hyperbolic links, in $SU(2)$ Chern-Simons theory. We showed that all torus links have a GHZ-like entanglement structure, and provided evidence to suggest that hyperbolic links tend to have a W-like entanglement structure, (iii) In order to get a better handle on hyperbolic links, we complexified the gauge group to $SL(2,\mathbb{C})$, where in the $\sigma \to \infty$ limit we were able to make partial analytical progress using known results from hyperbolic geometry on link complements. In particular, we showed that in the limit $\sigma \to \infty$, all hyperbolic links correspond to product states with no entanglement. Backing off from this limit, we observed that a certain matrix which appears in the Neumann-Zagier potential on the moduli-space of hyperbolic structures on the link complements controls the entanglement structure at leading order in $1/\sigma$. It would be interesting to use this last observation more fully.

There are several natural questions which present themselves at this stage. Does the SLOCC classification of entanglement structures from quantum information theory have a natural adaptation in knot theory to a classification of links? We saw a baby version of this idea manifest itself in the results of this paper, namely that all torus links have GHZ-like entanglement structures, while hyperbolic links seemingly have W-like entanglement structures. In other words, the GHZ/W-classification based on the robustness of the multi-party quantum entanglement seemingly translates to the torus/hyperbolic classification of links (although we should emphasize that we have not yet proved that all hyperbolic links are W-like). Further exploration is required to clarify whether SLOCC classification gives a useful way of characterizing links. A step in this direction would be to explore more detailed aspects of the entanglement structure of links. For instance, given an $n$-component link, we can assign to it a $(2^{n-1}-1)$-vector whose entries are the entanglement entropies of various bi-partitions of the link, a $3\times (\frac{1}{2}(3^{n-1}+1)-2^{n-1})$ matrix corresponding to the entanglement negativities of various tri-partitions, and so on. All these numbers can be computed directly from the colored Jones polynomial, and give a much more refined characterization of the entanglement structure of links. 

A second question is whether one can make useful progress in $SL(2,\mathbb{C})$ Chern-Simons theory by using the geometry of hyperbolic link complements. We have shown here that a certain matrix of coefficients in the Neumann-Zagier potential plays an important role. From a mathematical point of view then, it might be useful to study the properties of these coefficients in more detail for hyperbolic links. There is also a naive analogy one can make in this setup with the ``complexity = volume'' conjecture \cite{Stanford:2014jda}. There exists a state-integral model (see \cite{2007JGP....57.1895H, Dimofte:2009yn} for details), or in other words a tensor-network model, for constructing precisely the type of states we studied in the present paper for $SL(2,\mathbb{C})$. In these tensor-network models, one begins with the ideal-tetrahedral decomposition of the link complement (discussed in Appendix \ref{appHypStr}) and inserts one tensor per tetrahedron. The complexity $\mathcal{C}$ of such a network (i.e., the number of tensors in the full network) is naturally lower bounded by a constant times the hyperbolic volume of the link complement\footnote{This just follows from the trivial observation that the volume of an ideal hyperbolic tetrahedron is upper bounded by $\alpha^{-1} = 3 \Lambda(\pi/3)$, where $\Lambda(x)$ is the Lobachevsky function. } :
\beq
\mathcal{C} \geq \alpha\, V_{\mathrm{hyp}}.
\eeq
It would be interesting to see if one can carefully define the circuit complexity for these tensor networks and show that the ``optimal'' circuit (suitably defined) saturates this inequality.

From a holographic perspective, Chern-Simons theory is known to be dual to closed topological strings on resolved conifold geometries \cite{Gopakumar:1998ki}. It is clearly interesting to ask whether the entanglement entropy we have studied in this work has a suitable Ryu-Takayanagi interpretation from the closed string point of view. The bound on the minimal genus separating surfaces proved in this paper resembles the Ryu-Takayanagi minimal-area prescription (or more precisely the minimal-area bound which appears in MERA tensor networks), and might point to a deeper story underlying this resemblance. 

Finally, from a more practical viewpoint it is also an interesting question whether the entanglement we have studied in the present work has any applications to real materials. In particular, one wonders whether the states we have described can be constructed in the lab.

\subsection*{Acknowledgements}
We would like to thank Pawel Caputa, Ron Donagi, Nathan Dunfield, Sergei Gukov, Taylor Hughes, Mark Mezei, Eric Sharpe and Tadashi Takayanagi for useful conversations or email communications. We are particularly grateful to Tudor Dimofte and Alex Maloney for several useful conversations and communications,
 and to Alex Maloney for brief initial collaboration. Research funded by the Simons Foundation (\#385592, VB) 
 through
the It From Qubit Simons Collaboration, the US Department of Energy contract \#FG02-05ER-41367 and the US Department of Energy contract \#DE-SC0015655 (RGL).

\appendix

\section{Appendix: Hyperbolic geometry on Link Complements}\label{appHypStr}
In this Appendix, we spell out further details on how to construct the moduli space of hyperbolic structures on link complements and the attendant volume function, following \cite{Thurston, purcell2017hyperbolic, ratcliffe2006foundations, Dimofte:2009yn}. We first give a lightning summary for readers who do not wish to delve into the minutiae, which will then be followed by a detailed discussion.


The problem of interest is to construct hyperbolic structures on the link complement of a hyperbolic link. The most convenient way to do this is to build the link complement by gluing together a number of ideal tetrahedra in hyperbolic space. An ideal tetrahedron in $\mathbb{H}^3$ is a tetrahedron with all its vertices on the asymptotic boundary of $\mathbb{H}^3$. For instance if we take the half-space model of hyperbolic space with the metric
 \beq
 g_{\mathbb{H}^3} = \frac{dx_0^2 + dzd\bar{z}}{x_0^2},\;\;\;\cdots\;\;\;(z =x^1+ix^2)
 \eeq
  \begin{figure}[t]
 \centering
 \includegraphics[height=4cm]{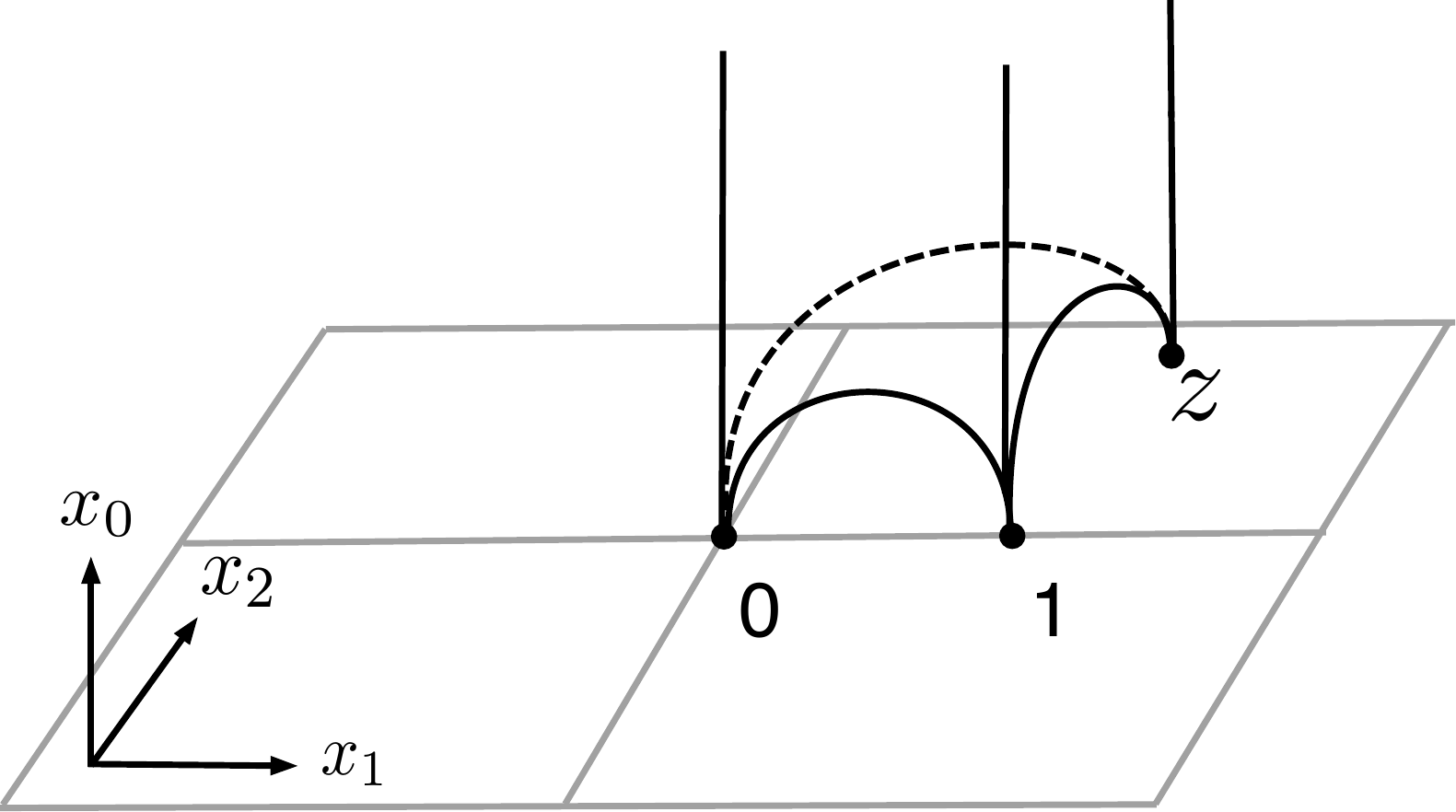}
 \caption{\textsf{An ideal tetrahedron in hyperbolic space with the shape parameter $z$ has all its vertices on the conformal boundary, three of them at $0, 1, z$ and the fourth vertex at $\infty$.}\label{tet}}
 \end{figure}
 then by conformal invariance we can choose three vertices of the tetrahedron to be at $0,\; 1$ and $\infty$ while the last vertex will be at $z \in \bC$, where all these points are understood to be on the conformal boundary of $\mathbb{H}^3$ (see Fig. \ref{tet}).  Thus, every ideal tetrahedron is labelled by one complex parameter $z$, which is often called the \emph{shape parameter}. The hyperbolic volume of such an ideal tetrahedron with shape parameter $z$ is given by 
 \beq\label{tetvol}
 \mathrm{Vol}(z) = \mathrm{Im}\Big(\mathrm{Li}_2(z)\Big) + \mathrm{arg}\,(1-z)\ln\,|z|.
 \eeq
Note that the volume is positive if $\mathrm{Im}\,z>0$, negative if $\mathrm{Im}\,z< 0$ (corresponding to opposite orientation) and zero if $z \in \mathbb{R}$ (corresponding to a degenerate tetrahedron). If the link complement is built out of $N$ tetrahedra, then we have $N$ independent complex variables, $\{z_n\}$ with $n=1,2,\cdots N$, to solve for. This is done as follows -- in gluing these tetrahedra to form the link complement, one must satisfy a list of algebraic conditions on the shape parameters. These conditions are of two types: (i) requiring consistent glueing at every edge (which lies in the interior of $S^3-N(\cL^n)$), namely the the sum of all the dihedral angles around the edge should be $2\pi$. These are often called \emph{edge-gluing} conditions, and are equivalent to requiring that the $SL(2,\bC)$ connection one is building is indeed flat everywhere in the bulk of the link complement. (ii) Requiring that the holonomies at the torus boundaries agree with the specified $m_i$'s (where recall $m_i =e^{u_i}$). These are called \emph{cusp} conditions. By solving the edge-gluing and cusp conditions together, one obtains solutions for the $\{z_n\}$s as functions of the $u_i$s. Generically, the solution is not unique, and one finds multiple branches of solutions which we will label by $\a$. These different branches should be interpreted as different saddle points in the path integral of Chern Simons theory.  The contribution to the path integral is simply
\beq \label{sp1}
e^{-\frac{\sigma}{\pi}V^{(\a)}(u_1,\cdots, u_n)}
\eeq
where $V^{(\a)} (u_1,\cdots,u_n)= \sum_{a=1}^n \mathrm{Vol}\,(z_a(u_1,\cdots,u_n))$. Note that in addition to being labelled by $\a$, the solutions are also parametrized by the continuous variables $u_i$; we therefore have \emph{moduli spaces} of flat connections (analogous to the Teichmuller spaces in the theory of Riemann surfaces) labelled by the coordinates $u_i$. 

We now present a more detailed review of the above construction. 
We will outline how to compute the volume function $V^{(\text{conj})}_M(u_1,u_2)$ for the Whitehead link complement $M = S^3 \setminus 5_1^2$.
\subsection*{Triangulation}
There exist algorithms which generate the link complement given only the link diagram.
However, we begin here directly from a visualization of the link complement.
The interested reader should see (\cite{purcell2017hyperbolic}) for an example of the link diagram-to-complement procedure.
\\\\
The Whitehead link complement may be drawn as an octahedron with a certain face gluing pattern (\cite{ratcliffe2006foundations}).
We will see that it is possible to put a hyperbolic structure on this manifold.
There are two vertices ($v$, $w$) and four edges ($a$, $b$, $c$, $d$).
\begin{figure}[h]
\centering
\includegraphics[scale=1]{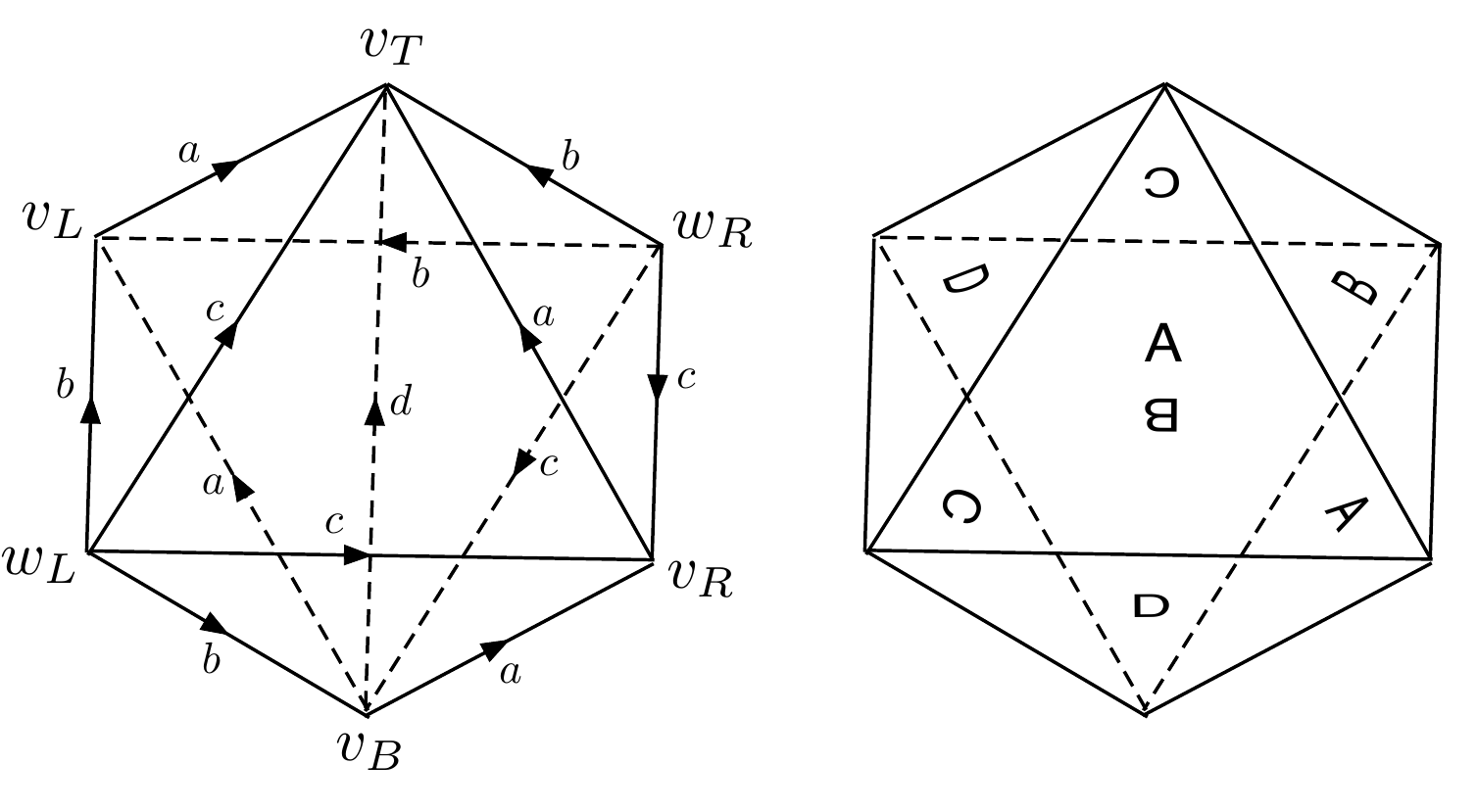}
\caption{\textsf{(Left) The vertex labels, edge labels, and orientations of the complement.  (Right) Face labels come in pairs, because faces with the same label are identified with each other.  Edge $d$ allows a breakdown of the octahedron into tetrahedron as seen in Fig.~\ref{decompfig}.}}
\label{octahedron}
\end{figure}
Keep in mind the vertices $v$ and $w$ are not actually part of the link complement; the ideal tetrahedra do not include their vertices.
We have labeled the vertices with subscripts to help visualize the decomposition into tetrahedra; remember that we really have $v_T = v_B = v_L = v_R$ and $w_L = w_R$.
\\\\
In order to compute meridianal holonomies, we require a visualization of the boundary torus before it pinches off into a cusp; we will informally refer to  such an image as the \textit{developing map} of a vertex.
To find the developing map for a vertex, in each tetrahedron we slice off all corners which contain that vertex, and use the face identifications to determine which newly created boundary edges are identified with each other.
The boundary faces (which are triangles, by construction) will then inherit a gluing from the tetrahedral face gluing, and will come together to form a torus triangulation.
We imagine the boundary torus shrinking down to a point, which corresponds to not slicing off any corners of ideal tetrahedra, to reconstruct the full 3-manifold $M$.
For the Whitehead link complement, we can proceed more easily by dealing with the octahedral form directly, and slicing off boundary squares as in Fig.~\ref{devmaps}.
Afterward, the square-tiled torus may be fully triangulated by inserting the edges associated with the additional faces created by insertion of edge $d$.
The two developing maps are shown in Fig.~\ref{devmaps}. 
\begin{figure}
\centering
\includegraphics[scale=.75]{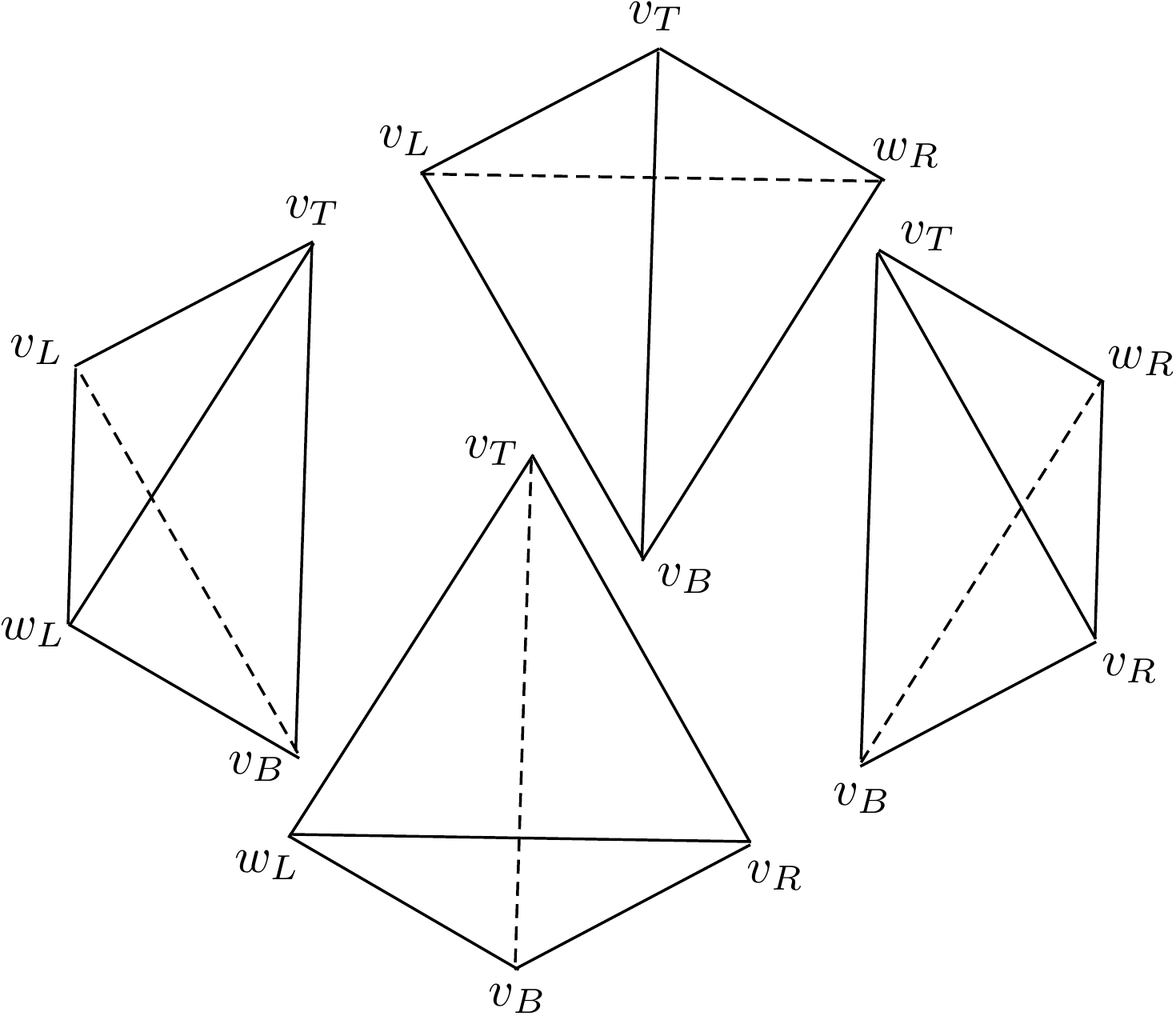}
\caption{\textsf{Decomposition of the Whitehead link complement into four ideal tetrahedra.  Orientations inherited from the octahedron in Fig.~\ref{octahedron} reveal two positively and two negatively oriented tetrahedra in the sense of Fig.~\ref{standard}.}}
\label{decompfig}
\end{figure}
\begin{figure}
\centering
\includegraphics[scale=.75]{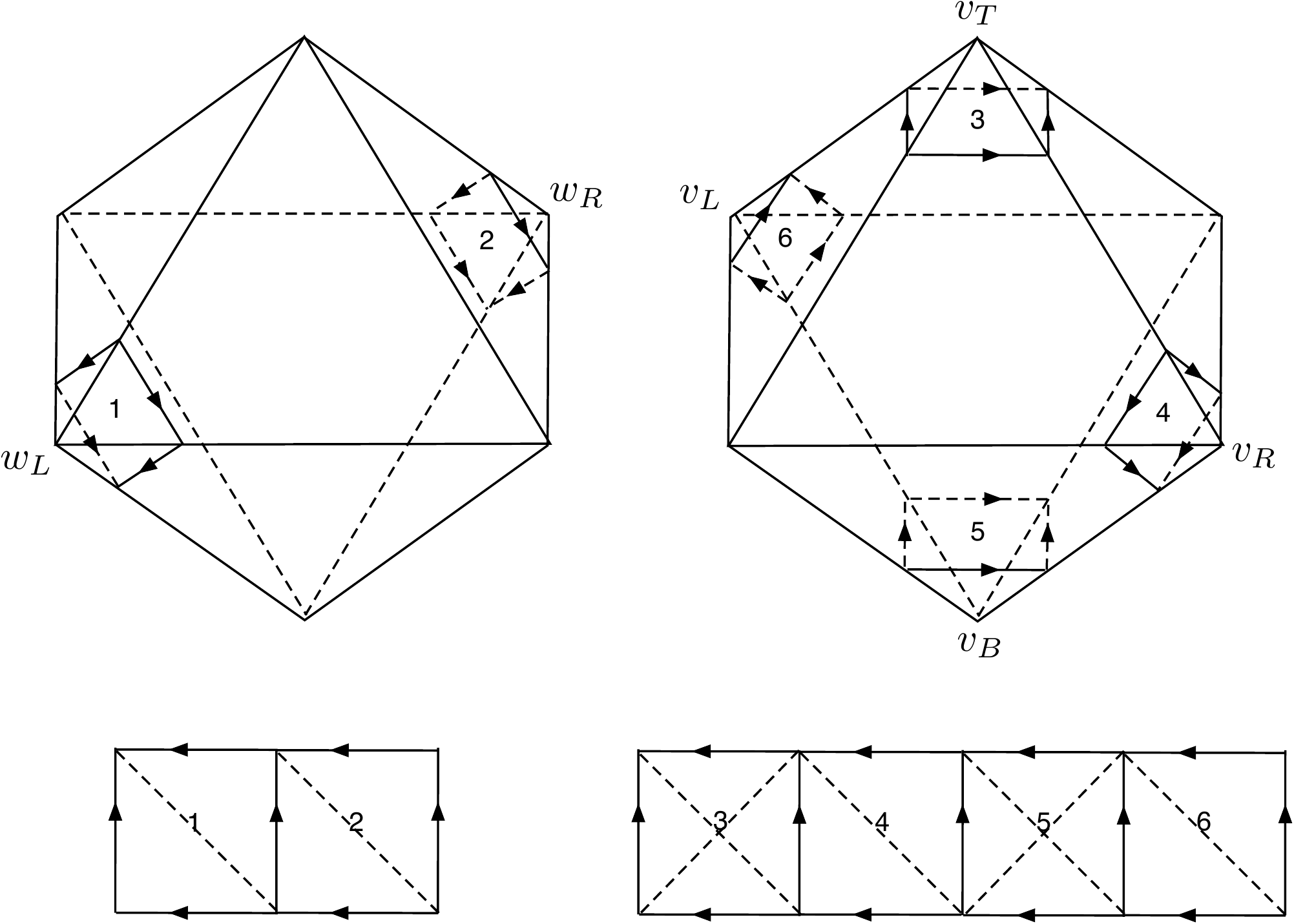}
\caption{\textsf{(Top) Slicing off squares for vertices $w$ (left) and $v$ (right) to generate a boundary tiling.  (Bottom) Developing maps for vertices $w$ (left) and $v$ (right).  The dotted lines in the developing maps represent edges generated by the addition of edge $d$ to the octahedron, i.e., the split into tetrahedra.}}
\label{devmaps}
\end{figure}
\subsection*{Hyperbolic Structure}
Oriented tetrahedral decomposition can be performed for any knot complement.
A complete hyperbolic structure, however, will only exist for hyperbolic links (like the Whitehead link).
Before constructing and solving the edge gluing and completeness equations, which will yield a moduli space of incomplete hyperbolic structures, we review some facts about embedding ideal tetrahedra in hyperbolic space.
\subsubsection*{Hyperbolic Tetrahedra}
Recall several facts about the upper half space model of 3-dimensional hyperbolic space $\mathbb{H}^3$.
We choose coordinates $(x,y,h)$ so that $\mathbb{H}^3 = \{ (x+i y, h) \in \mathbb{C} \times \mathbb{R}\ |\ h > 0 \}$.
Then, the metric is $ds^2 = h^{-2} (dx^2 + dy^2 + dh^2)$.
An ideal tetrahedron $\Delta$ embedded in $\mathbb{H}^3$ is a 3-simplex with all vertices lying on the boundary $\partial \mathbb{H}^3 = S^2$, and all edges lying on geodesics of $\mathbb{H}^3$.
Note that the point at infinity makes $\partial \mathbb{H}^3$ a plane ($\mathbb{C}$) plus a point, which by stereographic projection is topologically a two-sphere.
Geodesics in $\mathbb{H}^3$ are given by lines and semicircles that intersect $\partial \mathbb{H}^3$ perpendicularly.
Using an isometry of $\mathbb{H}^3$, we can send three vertices of the ideal tetrahedron to the points 0, 1, and $\infty$.
The fourth vertex lies at a point $z \in \mathbb{C}$, called the \textit{shape parameter} of $\Delta$.
The shape parameter contains complete information about all dihedral angles, edge lengths, and even the hyperbolic volume contained in $\Delta$, and a generic ideal tetrahedron in $\mathbb{H}^3$ may be labeled $\Delta(z)$.
It follows that the dihedral angle associated to any edge can be encoded in the argument of one of three complex quantities, which are all functions of the shape parameter.
These three quantities are called \textit{edge parameters}, and are denoted
\begin{equation}
z_1 = z, \hspace{1cm} z_2 = \frac{1}{1-z}, \hspace{1cm} z_3 = 1-\frac{1}{z}.
\end{equation}
Before returning to the Whitehead link, we draw attention to the fact that despite having edges of infinite length, $\Delta(z)$ has a finite hyperbolic volume.
\subsubsection*{Edge Gluing and Completeness Equations}
The edge gluing equations are obtained by taking the product over edge parameters assigned to all instances of a particular edge in the tetrahedra (\cite{purcell2017hyperbolic}), and setting the result equal to one.
Recall from the previous discussion that the edge parameters encode dihedral angles in their arguments, so this is exactly the condition that the sum of angles around a given edge is $2\pi$, which prevents the emergence of any angular deficits at points along the edge.
\begin{figure}
\centering
\includegraphics[scale=.75]{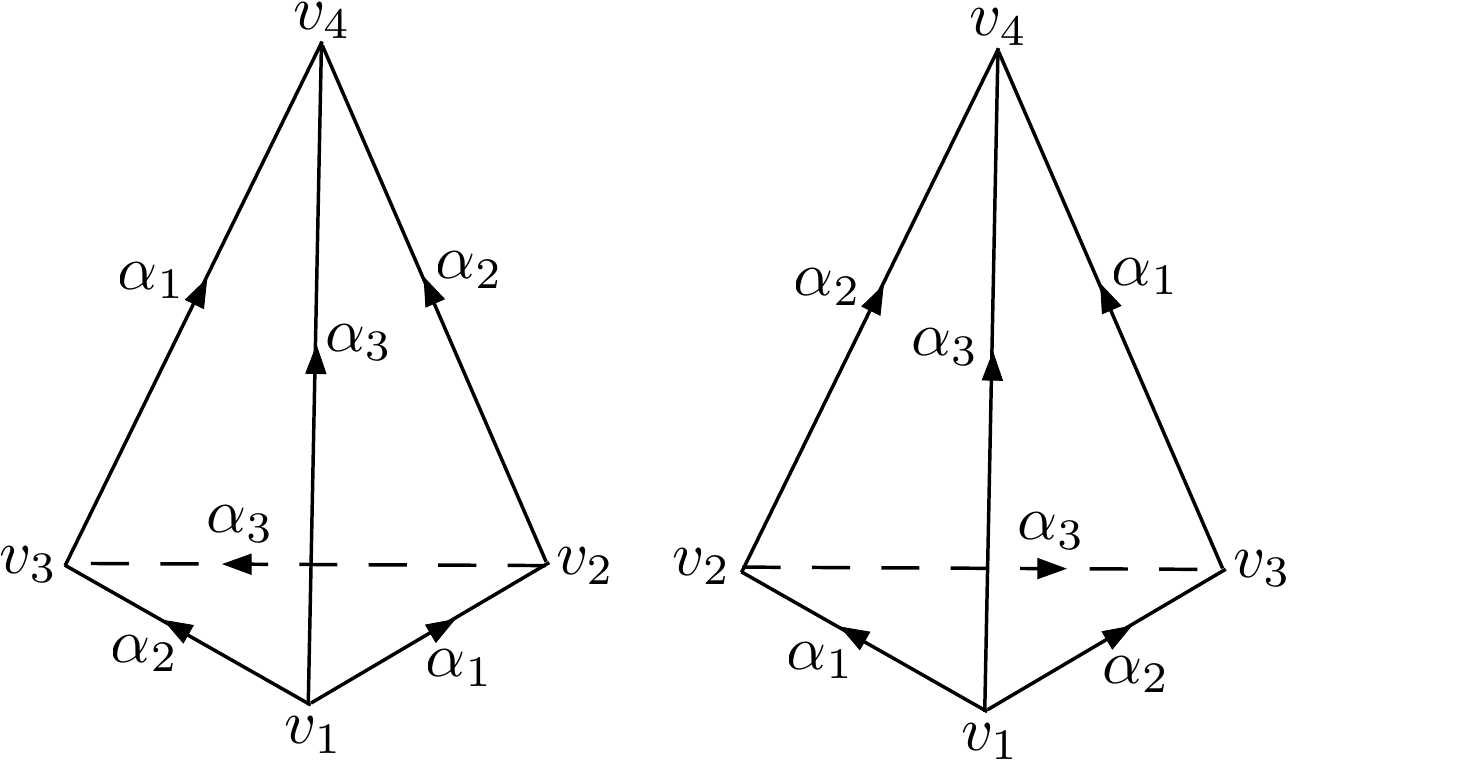}
\caption{\textsf{Positively (left) and negatively (right) oriented tetrahedra with shape parameter $\alpha$.  The Whitehead link complement has two of each orientation, for a total of four tetrahedra.}}
\label{standard}
\end{figure}
In particular, the Whitehead link complement splits into two positively and two negatively oriented tetrahedra in the sense of Fig.~\ref{standard}.
Label the shape parameters $w$, $x$, $y$, and $z$, corresponding to the tetrahedron with face $A$, $B$, $C$, and $D$, respectively, as denoted in Fig.~\ref{octahedron}.
In the two standard tetrahedron orientations, edge parameters for $\Delta(\alpha)$ are assigned by $[v_1, v_2] \to \alpha_1$, $[v_1,v_3] \to \alpha_2$, $[v_2,v_3] \to \alpha_3$, and then adding the only missing edge parameter to the remaining edge on each vertex, i.e., $[v_1,v_4] \to \alpha_3$, $[v_2,v_4] \to \alpha_2$, and $[v_3,v_4] \to \alpha_1$.
These edge parameters are shown next to their corresponding edges in Fig.~\ref{standard}.
The four edges of the Whitehead link complement then yield the following four edge gluing equations (listed in alphabetical order from $a$ to $d$).
\begin{equation}
\begin{split}
1 & = w_2 w_3 x_1 x_3 y_2 y_3 z_1 z_3 \\
1 & = w_2 x_3 y_1 y_3 z_1 z_2 \\
1 & = w_1 w_3 x_1 x_2 y_2 z_3 \\
1 & = w_1 x_2 y_1 z_2
\end{split}
\label{gluing}
\end{equation}
The completeness equations are slightly more subtle.
Note that any vertex of a triangle in a developing map is associated with an edge parameter, which is precisely the edge parameter of the tetrahedron edge that is intersecting that vertex.
For a more detailed representation of this, see (\cite{ratcliffe2006foundations}).
Now observe that any meridian may be deformed in such a way as to slice off a single corner from every triangle through which it passes.
The completeness relations are computed by setting 1 equal to the product of the edge parameters of corners to the left of the meridian and inverse edge parameters of corners to the right.
However, we want to allow for incomplete hyperbolic structures as well, which correspond to setting these products equal to a complex number $m = e^{u}$.
Since there are two boundary tori, we have two completeness relations for complex numbers $m_1$ and $m_2$.
\begin{equation}
\begin{split}
y_2 x_3^{-1} = m_1 \\
x_3 z_3^{-1} = m_2
\end{split}
\label{completeness}
\end{equation}
This system is not unique; for another example of an equivalent Whitehead link complement system, see (\cite{apanasov1992topology}).
We want to solve the system formed by equations (\ref{completeness}) and (\ref{gluing}) for the shape parameters, so we can use them to compute the hyperbolic volume of $M$ by adding the volume of the individual tetrahedra in the decomposition.
At first, it seems like we have six equations and only four unknowns, so the system is  overdetermined.
However, it turns out that two of the gluing equations are redundant; this is one of  several coincidences that must occur for a link complement to admit a complete hyperbolic structure.
There are three solutions; two of them are the geometric and geometric conjugate branches of the A-polynomial, which correspond to the flat connections that achieve maximum and minimum volume, respectively, at $u_1 = u_2 = 0$.
In other words, we chose the geometric conjugate branch as the one that reaches minimum volume at $m_1 \equiv e^{u_1} = 1$ and $m_2 \equiv e^{u_2} = 1$.
We now turn to the volume formula itself.
\subsection*{Hyperbolic Volume}
The volume of a hyperbolic 3-manifold $N$ with $k$ boundary tori that has been decomposed into $n$ ideal tetrahedra with shape parameters $\alpha_i$ is
\begin{equation}
V^{(\beta)}_N(u_1, \dots , u_k) = \sum_{i=1}^n D \left[ \alpha_i(u_1, \dots, u_k) \right]
\end{equation}
where $u_j$ is the holonomy eigenvalue $m_j \equiv e^{u_j}$ of the $j^{\text{th}}$ boundary torus, and $\beta$ labels the solution of (\ref{gluing}) and (\ref{completeness}) we have chosen.
$D(\alpha)$ is the Bloch-Wigner function, defined as
\begin{equation}
D(\alpha) \equiv \text{Im}(\text{Li}_2(\alpha)) + \arg (1-\alpha) \log |\alpha|
\end{equation}
where $\text{Li}_2$ is the dilogarithm and $\arg$ returns the angle $\theta \in (-\pi,\pi]$ that its argument makes with the real axis in the complex plane.
At the volume minimum on the conjugate branch, i.e., $m_1 = m_2 = 1$, our Whitehead link complement shape parameters become
\begin{equation}
w = y = -i, \hspace{1cm} x = z = 1-i
\end{equation}
The hyperbolic volume of $M$ on the conjugate branch at the saddle point is therefore
\begin{equation}
V^{(\text{conj})}_M(u_1 = 0,u_2 = 0) = 2 D\left( -i \right) + 2 D(1-i) \approx -3.664
\end{equation}
Note that the parameters $u_j$ are complex, and so admit a decomposition as $u_{j1} + i u_{j2}$, and we write the full set of holonomy eigenvalues as $u_{jk}$ with $k \in \{1,2\}$.
This allows us to write the volume $V_M$ as a function of four real variables as opposed to two complex variables.
Also, let $\alpha_j^R$ and $\alpha_j^I$ be the real and imaginary parts of the shape parameter $\alpha_j$.
The Bloch-Wigner function in this form is
\begin{equation}
D(\alpha^R, \alpha^I) = \text{Im} (\text{Li}_2(\alpha^R + i \alpha^I)) + \arctan \left( 1-\alpha^R, -\alpha^I \right) \log \left[ \sqrt{(\alpha^R)^2 + (\alpha^I)^2} \right]
\end{equation}
where the arctangent function is defined with two variables to give the angle in the appropriate quadrant.
It may be expressed as a piecewise function of the usual one-variable arctangent.
Using this formula together with the solution of (\ref{completeness}) and (\ref{gluing}) yields the full (manifestly non-holomorphic) volume function $V_M^{\text{(conj)}}(u_{1x},u_{1y},u_{2x},u_{2y})$ on the conjugate branch of the moduli space of hyperbolic structures.
\subsection*{Calculating the quartic coefficients}\label{sect:AppQuart}
Possessing the gluing and completeness equations of the triangularization of a particular hyperbolic link, we can then compute the ingredients of the link state \eqref{sp3}.  As outlined in Section \ref{sec5}, the quantum entanglement of link state appears at order $1/\sigma$ as $\sigma\rightarrow\infty$ which is the quartic term, $A_{ij}$, in the expansion of the hyperbolic volume, \eqref{eq:Vexpand}.  Writing explicitly, $u_j=u_j^R+i\,u_j^I$ and $A_{ij}=A^R_{ij}+i\,A^I_{ij}$ we have at quartic order
\begin{equation}
V^{(4)}=\frac{1}{\sigma^2}\sum_{i,j}\left\{A^R_{ij}u_i^Iu_i^R(u_j^R)^2-\frac{1}{8}A_{ij}^I\left(4u_i^Iu_i^Ru_j^Iu_j^R-3(u_i^R)^2(u_j^R)^2+2(u_i^R)^2(u_j^I)^2+(u_i^I)^2(u_j^I)^2\right)\right\}.
\end{equation}
Therefore we can extract the real and imaginary parts of $A_{ij}$ separately by taking appropriate derivatives at the saddle point:
\begin{align}\label{eq:d4V}
\sigma^2\left.\frac{\pa^4V}{(\pa u_i^R)^2(\pa u_j^R)^2}\right|_{u_i=0}=&3A_{ij}^I+6 A_{ii}^I\delta_{ij}\nonumber\\
\sigma^2\left.\frac{\pa^4V}{\pa u_i^I\pa u_i^R(\pa u_j^R)^2}\right|_{u_i=0}=&2 A^R_{ij}+4 A^R_{ii}\,\delta_{ij}.
\end{align}
A straight forward approach would be to solve the gluing equations at generic $u_i$ for shape parameters $z^\alpha(u_i)$ and then perform the chain rule.  In practice, this can be a cumbersome calculation especially for links with three or more components or triangulations with many shape parameters.  Fortunately we can circumvent this difficulty by differentiating the gluing and completeness equations directly.  In this approach the computation of \eqref{eq:d4V} is quite simple.  For instance, differentiating \eqref{completeness}
\begin{align}\label{dercompleteness}
\frac{\pa y_2}{\pa u_i^R}x_3^{-1}-y_2\,x_3^{-2}\frac{\pa x_3}{\pa u_i^R}=\delta_{1i}\nonumber\\
\frac{\pa x_3}{\pa u_i^R}z_3^{-1}-x_3\,z_3^{-2}\frac{\pa z_3}{\pa u_i^R}=\delta_{2i}
\end{align}
Evaluating the above expression at the saddle point, $u_i=0$ (and doing this for the gluing equations, \eqref{gluing}) gives us an expression for the first derivatives in terms of the values of the shape parameters at the $u_i=0$ point.  Taking another derivative of \eqref{dercompleteness} we can repeat this and solve for the second derivatives in terms of the first and so on.  Performing the chain rule we can then, at least in principle, calculate $A_{ij}$ directly from the gluing/completeness equations.\\
\\
We have walked through the derivation of the gluing and completeness equations for the Whitehead link as a pedagogical exercise; for more general links we can take advantage of SnapPy which has the gluing/completeness equations catalogued.  This makes the calculation of $A_{ij}$ rather efficient using a computer algebra system.  As an example, pulling the SnapPy data for the link L6a4 (the Borromean rings) we were able to compute numerically\footnote{up to an error of order $\pm 10^{-15}$}:
\begin{align}
\left.\sigma^2\frac{\pa^4 V}{(\pa u_i^R)^2(\pa u_j^R)^2}\right|_{u_i=0}&=192\left(\begin{array}{ccc}-1&1&1\\1&-1&1\\1&1&-1\end{array}\right)\nonumber\\
\left.\sigma^2\frac{\pa^4 V}{\pa u_i^I\pa u_i^R(\pa u_j^R)^2}\right|_{u_i=0}&=0
\end{align}
giving \eqref{eq:ABorr}.

\providecommand{\href}[2]{#2}\begingroup\raggedright\endgroup

\end{document}